\documentclass[aps,twocolumn,showpacs,superscriptaddress,amsmath,amssymb]{revtex4}
\usepackage{graphics,epsf,colordvi}

\newcommand{\bea}{\begin{eqnarray}}
\newcommand{\beq}{\begin{equation}}
\newcommand{\eea}{\end{eqnarray}}
\newcommand{\eeq}{\end{equation}}
\begin{document}
\title{
Magnetic alteration of entanglement in two-electron quantum dots}
\author{N. S. Simonovi\'c}
\affiliation{Institute of Physics, University of Belgrade, P.O.
Box 57, 11001 Belgrade, Serbia}
\author{R.G. Nazmitdinov}
\affiliation{Departament de F{\'\i}sica,
Universitat de les Illes Balears, E-07122 Palma de Mallorca, Spain}
\affiliation{Bogoliubov Laboratory of Theoretical Physics,
Joint Institute for Nuclear Research, 141980 Dubna, Russia}

\begin{abstract}
Quantum entanglement is analyzed thoroughly in the case of
the ground and lowest states of two-electron axially symmetric
quantum dots under a perpendicular magnetic field. The
individual-particle and the center-of-mass representations are
used to study the entanglement variation at the
transition from interacting to noninteracting particle regimes.
The mechanism of symmetry breaking due to the interaction, that
results in the states with symmetries related to the later
representation only, being entangled even at the vanishing
interaction, is discussed. The analytical expression for the
entanglement measure based on the linear entropy is derived in the
limit of noninteracting electrons. It reproduces remarkably well
the numerical results for the lowest states with the magnetic
quantum number $M \ge 2$ in the interacting regime. It is found
that the entanglement of the ground state is a discontinuous
function of the field strength. A method to estimate the
entanglement  of the ground state, characterized by the quantum
number $M$, with the aid of the magnetic field dependence of the
addition energy is proposed.
\end{abstract}
\maketitle

\section{Introduction}

Quantum entanglement and its
implications  such as quantum teleportation, quantum
cryptography, and quantum computation, to name just a few
(for a review see \cite{nielsen,bengtsson,barn}),
are the subject of intense research efforts in recent
years.  Apart from possible practical
applications, these research lines provide a deeper
understanding of the fundamental aspects of quantum correlations
in many-body systems. This issue
is currently the focus of an increasing research activity,
establishing new connections between quantum information theory and
quantum correlations in atomic, molecular and condensed
matter physics \cite{r2,buc}.

Among suitable systems, capable of demonstrating
a relationship between quantum entanglement  and quantum correlations, are quantum dots (QDs).
They offer one of the perspective experimental platforms
for quantum communications in  solid-state environment \cite{bul}
and allow the study of various aspects of quantum correlations
with a high accuracy \cite{man,rev1,sar}.
Indeed, control of QD spectra by gates and an external magnetic field
provides an efficient way to create a single-electron ground state manifold with
well-defined spin states. According to a general wisdom, spin degrees of freedom
in QDs are promising candidates for quantum information processing
(see, e.g., \cite{vin,han}). At the same time, quantum communication research considers
photons as flying qubits that carry quantum information over long distances.
QDs appear as remarkable candidates to realize the
interface between stationary qubits and flying photonic qubits.
The optical excitation spectra of QDs exhibits strong
correlations between the initial electron state and light polarization.
This leads to the possibility to explore QD spin-photon entanglement, that
nowadays emerges in the fast developing field of coherent optical manipulation
of spin states in a solid-state environment (for a review see \cite{gao}).

To operate efficiently by stationary qubits it is necessary to
deepen our understanding of
various aspects of quantum correlations  brought about
by the electron-electron interaction, an effective confining potential, external fields, and
the degree of entanglement in QDs.
Besides computational problems for many-body quantum systems,
one has to address to the problem of measuring
entanglement for indistinguishable particles.
In other words, one has to discriminate entanglement from correlations due to the statistics of
indistinguishable particles \cite{schl,ber,NV}.
In spite of this difficulty, bipartite entanglement has been investigated
in a number of systems of physical interest with the aid of various numerical approaches
at zero magnetic field.
In particular, the connection between entanglement and correlation energy has been studied
in the context of a two-electron artificial atom \cite{mar}.
The amount of entanglement of the ground state and of the first few excited states of helium
was assessed by using high-quality state-of-the-art wavefunctions \cite{pl1}.
Benenti et al. \cite{ben} used a configuration-interaction variational method
to compute the von Neumann and the linear entropy for several low-energy singlet
and triplet eigenstates of helium.
Lin et al. \cite{lin} computed the spatial quantum entanglement
of helium-like ions by means of spline techniques.
Various types of wave functions have been used to increase
the numerical accuracy of  the entanglement measures
of helium-like atoms with a few electrons \cite{h1,h2,rosa}.
Also, the entanglement properties of bound states in the two-electron
Moshinsky model \cite{YPD} and a solvable model of two-electron
parabolic quantum dot \cite{ ch1} were investigated.

Two-electron QDs being realistic non-trivial systems are
particularly attractive for a detailed analysis of quantum
correlations (for a review see  \cite{rev2}). In contrast to many electron QDs, their eigenstates
can be obtained very accurately, or in some cases exactly. The
same conveniences also hold while calculating appropriate
entanglement measures.
The spatial entanglement measures have been used to test the validity
of approximations of the density-functional theory for
Hook's atom \cite{sud}.
It represents a possible model for describing
two electrons trapped in a quantum dot.
Further, it was found that the entanglement of the singlet state increases
with the increase of the dimension of the Hook's system \cite{K2}.
It was shown that the anisotropy of the confining potential
of  two-dimensional two-electron QD
drastically influences the entanglement properties in
a strong correlation regime \cite{K1}.
The entanglement (linear entropy) has been calculated to
trace the transition from bound to continuum states in
two-electron QDs \cite{rs}.
The study of a two-electron QD
by means of entanglement witness provided new aspects
in the ongoing discussion about the origin of Hund's rules in atoms \cite{frid}.
It was shown that the presence of the donor-impurity
has  little   impact on the entanglement in two-electron QDs
in the regime of formation of the Wigner molecule \cite{K3}.

We recall that all these studies are mostly focused
on the  interconnection between quantum correlations and
various measures of entanglement at zero
magnetic field, with a different degree of numerical accuracy.
In this paper we will analyze this relationship in
the lowest states of two-electron QDs, that are evolving to the
ground states in different intervals of the magnetic field
strength. To this aim we will present a numerical and an analytical approach that
describes remarkably well the numerical results.
Special attention is paid to the study of quantum correlations
and the entanglement in the limiting case of noninteracting
electrons in order to elucidate the impact of the electron-electron
interaction and/or  the magnetic field.
Preliminary results of our analysis have been presented
in \cite{NSPC,NS13}, where the influence of the magnetic field on
the entanglement of the singlet $m=0$ states only has been investigated.

The structure of the paper is as follows.
In Sec.~\ref{sec:model-phyq-st} we set up the model for axially
symmetric two-electron QDs in a magnetic field directed along the
symmetry axis, analyze integrals of motion and symmetries in the
cases with and without the electron-electron ($V_C$) interaction,
and introduce appropriate basis sets. We also consider here
entanglement measures for two-electron states. In
Secs.~\ref{sec:indiv_pd} and \ref{sec:CM-rep}, we analyze the
orbital properties of the lowest eigenstates in the limit $V_C \to
0$, in the individual particle and the
center-of-mass representations, respectively. In
Sec.~\ref{sec:CM-rep} we consider also the dependence of the
entanglement of QD lowest states on the strengths of the Coulomb
interaction between electrons as well as on the magnetic field.
Sec.~\ref{sec:CM-Rel-rep} establishes the transformation of one
representation to another one. The entanglement of the ground
state, particularly in the light of the so-called singlet-triplet
transitions, is analyzed in Sec.~\ref{sec:ground_st}. In
Sec.~\ref{exme} we describe how to estimate the entanglement
at different field strengths by means of experimental
observations. Main conclusions are summarized in
Sec.~\ref{sec:conclusions}. In Appendices A--E we present some
technical details of our analysis.

\section{The model}
\label{sec:model-phyq-st}
\subsection{The Hamiltonian}
Our  analysis is based on the  Hamiltonian
\begin{equation}
H_\mathrm{tot} = \sum_{i=1}^{2}H_i + V_C+H_S=H+H_S,
\label{hamtot}
\end{equation}
where
\begin{equation}
H_i = \frac{1}{2m^*\!}\,(\mathbf{p}_i - e\mathbf{A}_i)^{\! 2} +
U(\mathbf{r}_i),
\end{equation}
The term $V_C = \lambda/\vert \mathbf{r}_1\!-\!\mathbf{r}_2\vert$
with $\lambda = e^2/4\pi\varepsilon_0\varepsilon_r$ describes the
Coulomb interaction between electrons. The constants $m^*$, $e$,
$\varepsilon_0$ and $\varepsilon_r$ are the effective electron
mass, unit charge, vacuum and relative dielectric constants of a
semiconductor, respectively.  In the expressions above
$\mathbf{r}_i$ and $\mathbf{p}_i$ ($i = 1,2$) are the positions
and momenta of the electrons, respectively, and $U(\mathbf{r}_i)$ is
the $i$-th electron confining potential. The orbital and spin
degrees of freedom (the terms $H$ and $H_S$ in
Eq.~(\ref{hamtot})) are fully decoupled in our model, since we
neglect the spin-orbit interaction

The influence of the magnetic field (we set the $z$-axis along the
vector $\mathbf{B}$) on the electron orbital motion is introduced
through the vector potential with gauge $\mathbf{A} = \frac{1}{2}
\mathbf{B} \times \mathbf{r} = \frac{1}{2}B(-y,x,0)$. The term
$H_S =g^* \mu_B B S_z/\hbar$ in (\ref{hamtot}) describes the
interaction of the (total) spin $\mathbf{S} = \mathbf{s}_1 +
\mathbf{s}_2$ with the magnetic field. Here $S_z$ is the
$z$-projection of the total spin, $g^*$ is the effective Land\'e
facto, and $\mu_B=|e|\hbar/2m_e$ is the Bohr magneton.

For axially symmetric QDs (with  $z$-symmetry axis)
the confining potential is quite well approximated
by the parabolic model \cite{rev2}. We consider the form
$U(\mathbf{r}) = \frac{1}{2}m^* u(\omega_0;\mathbf{r})$, where
\begin{equation}
u(\omega_0;\mathbf{r}) = \omega_0^2(x^2 + y^2) + \omega_z^2 z^2.
\label{confpot}
\end{equation}
Here $\hbar\omega_0$ and $\hbar\omega_z$ are the energy scales of
the confinement in the $xy$-plane (lateral confinement) and in the
$z$-direction (vertical confinement), respectively. We also introduce
the characteristic lengths of the lateral and vertical
confinements, $\ell_0 = \sqrt{\hbar/m^*\omega_0}$ and $\ell_z =
\sqrt{\hbar/m^*\omega_z}$, respectively.
In the extremely anisotropic case $\omega_z \gg \omega_0$ one has
$\langle z^2 \rangle \ll \langle x^2 + y^2 \rangle$, and the
Coulomb term $V_C$ reduces to the form $\lambda/[(x_1-x_2)^2 +
(y_1-y_2)^2]^{1/2}$. As a consequence, the motions in the
$xy$-plane and along the $z$-axis are practically decoupled (see
discussion in \cite{effch}). In this case the 2D (planar) model is
a good approximation.  This is due to the fact that electrons
perform only fast harmonic oscillations in the $z$-direction. As a
result, the states with the lowest energy of $z$-component
($2\times\hbar\omega_z/2$) are occupied only.

Applying the above given gauge condition, we have for
the single-electron Hamiltonians
\begin{equation}
H_i = \frac{\mathbf{p}_i^2}{2m^*} + \frac{1}{2}m^*
u(\Omega;\mathbf{r}_i) - \omega_L l_z^{(i)}, \label{hamsing}
\end{equation}
where $\omega_L = eB/2m^*$ is the Larmor frequency and $\Omega =
(\omega_0^2 + \omega_L^2)^{1/2}$ is the effective lateral
confinement frequency that depends through the frequency
$\omega_{L}$ on the magnetic field. The operator $l_z^{(i)}$ is
the $z$-projection of the angular momentum of the $i$-th electron.
Evidently, in the approximation of noninteracting electrons ($V_C
= 0$) the single-electron energies and orbital momenta are
integrals of motion. At $V_C\neq 0$ the quantities related to the
electron collective dynamics are conserved only (see below).

By introducing the center of mass (c.m.) $\mathbf{R}
=\frac{1}{2}(\mathbf{r}_1 + \mathbf{r}_2)$ and relative
$\mathbf{r}_{12} = \mathbf{r}_1 - \mathbf{r}_2$ coordinates,
the orbital part of Hamiltonian (\ref{hamtot}) separates into the
c.m. and relative motion terms, $H = H_{\rm c.m.} + H_{\rm rel}$ (see
details in \cite{rev2}), in agreement with the Kohn theorem
\cite{kohn}. The c.m. term has the form
\begin{equation}
H_\mathrm{c.m.} = \frac{{\mathbf P}^2}{2{\cal M}} + \frac{1}{2}{\cal
M} u(\Omega;\mathbf{R}) - \omega_L l^{(c.m.)}_z,
\label{cmham}
\end{equation}
where ${\cal M} = 2m^*$ is the total mass, and $l^{(c.m.)}_z$
is the $z$-projection of the c.m. angular momentum. The relative
motion term includes the Coulomb interaction between electrons
\begin{equation}
H_{\rm rel} = \frac{{\mathbf p}_{12}^2}{2\mu} + \frac{1}{2} \mu
u(\Omega;\mathbf{r}_{12}) - \omega_L l^\mathrm{(rel)}_z +
\frac{\lambda}{r_{12}} = H_{\rm rel}^{(0)} + V_C. \label{relham}
\end{equation}
Here $\mu = m^*/2$ is the reduced mass, and $l^\mathrm{(rel)}_z$
is the $z$-projection of the angular momentum for the relative
motion. Both projections  $l^{(c.m.)}_z$,
$l^\mathrm{(rel)}_z$, are integrals of motion due to the axial
symmetry of the system.

Finally, it should be mentioned that the Hamiltonian $H$ and the
corresponding equations of motion are invariant under the scaling
transformations $\mathbf{r}_i \to \mathbf{r}_i/\ell_0$,
$\mathbf{p}_i \to \mathbf{p}_i\ell_0/\hbar$ and $E \to
E/\hbar\omega_0$. As a consequence, the Hamiltonian $H$ expressed
in scaled variables is invariant under the simultaneous
variations of the QD parameters that keep constant the ratio
$\omega_z/\omega_0$ and the scaled strength of the Coulomb
interaction $\lambda/(\hbar\omega_0\ell_0) \equiv \ell_0/a^* =
R_W$ (the so-called Wigner parameter). Here, $a^* =
\hbar^2/\lambda m^*$ is the effective Bohr radius. In other words,
the whole class of axially symmetric QDs can be covered by varying
only these two parameters, and the additional parameter
$\omega_L/\omega_0$ if $B\neq0$. In particular, we will use the
value $R_W = 2$ that corresponds to a typical GaAs QD ($m^*
= 0.067\, m_e$, $\varepsilon_r \approx 12$, $\hbar\omega_0 =
3.165$\,meV.)

\subsection{Integrals of motion and symmetries}
\label{sec:int-of-motion}

As discussed above, the orbital and the spin degrees of freedom
are decoupled in our system. As a result, the eigenstates of
Hamiltonian (\ref{hamtot}) take the form
\begin{equation}
\label{psitot} \Psi = \psi(\mathbf{r}_1,\mathbf{r}_2)
\chi(\sigma_1,\sigma_2)\,,
\end{equation}
where $\psi$ and $\chi$ are the orbital and spin parts,
respectively. The spins of individual electrons and their
$z$-projections as well as the total spin and its $z$-projection
are the integrals of motion. We have two alternative sets of spin
quantum numbers: $\{s_1$, $s_2$, $m_{s1}$, $m_{s2}\}$ and $\{s_1$,
$s_2$, $S$, $M_S\}$. Here we choose the second set that provides a
definite exchange symmetry of two-electron states.

The orbital motion is characterized by several integrals of
motion. The corresponding operators $H_\mathrm{c.m.}$,
$l_z^{(c.m.)}$, $H_\mathrm{rel}$, $l_z^\mathrm{(rel)}$ (and
$H_z^{(c.m.)}$ in the 3D case) commute with $H$, and  also
mutually. Since in the 2D case there are four independent
integrals of motion that are in involution, the 2D model (it has
four orbital degrees of freedom) is fully integrable. Therefore,
these four operators define a complete set of commuting
observables (CSCO). In the 3D case the system is, however,
generally non-integrable at $V_C\neq0$, excluding some special
cases \cite{sc,hidsym,hd2}. Evidently, we have to use  as basis
states those that are defined at $V_C = 0$.

For noninteracting electrons there are two appropriate sets of
independent and commuting observables: (i) $\{H_1$,
$l_z^{(1)}$, $(H_z^{(1)})$, $H_2$, $l_z^{(2)}$, $(H_z^{(2)}) \}$
and (ii) $\{H_\mathrm{c.m.}$, $l_z^{(c.m.)}$,
$(H_z^{(c.m.)})$, $H_\mathrm{rel}^{(0)}$,
$l_z^\mathrm{(rel)}$, $(H_{z(0)}^\mathrm{(rel)})\}$.
Simultaneously, all these observables commute with $H$  at $V_C =
0$. Note, however, that the observables from different sets are
not commuting. For example,
\begin{eqnarray}
[l_z^{(1)},l_z^{(c.m.)}] \!\!&=&\!\!
-[l_z^{(2)},l_z^{(c.m.)}] \nonumber
\\
\!\!&=&\!\! \frac{i\hbar}{4} (x_1 p_x^{(2)} \!+\! y_1 p_y^{(2)}
\!-\! x_2 p_x^{(1)} \!-\! y_2 p_y^{(1)}),
\end{eqnarray}
etc. It appears that the sets (i) and (ii) are two alternative
CSCOs.

We recall a well known theorem: {\it If there are two conserved
physical quantities whose operators do not commute, then the
energy levels of the system are in general degenerate} (see \S 10
in Ref.~\onlinecite{LL}). Evidently, the existence of two sets at
$V_C=0$ yields the degeneracy of the eigenenergies in our system.
This degeneracy is easy to understand, taking into account  two
decompositions
\begin{equation}
H_0\equiv H(V_C=0) = \left\{\!\!
\begin{array}{l}
H_1 + H_2,
\\
H_\mathrm{c.m.} + H_\mathrm{rel}^{(0)}.
\end{array}
\right. \label{separation}
\end{equation}
The choice  of the first or the second decomposition
defines  the use of the set (i) or the set (ii), respectively.

The interaction $V_C$ breaks the symmetries of the noninteracting
model related to the set (i). As a consequence, it removes the
degeneracy existing in the noninteracting case. Note that the
symmetries of set (ii) are, however,  preserved.  Below we
will see that these states are generally entangled, even at $V_C
\to 0$; the removal  of the degeneracy by the inter-particle
interaction is the key point to understanding this
feature.

\subsection{Representations}

At $V_C=0$ the integrals of motion suggest two appropriate basis sets
for the orbital eigenfunctions $\psi(\mathbf{r}_1, \mathbf{r}_2)$ --
the individual-particle (IP) basis with
the eigenstates of the CSCO (i), and the
center-of-mass--relative-motion basis with the eigenstates of
the CSCO (ii). Hereafter, for the sake of convenience we name
the center-of-mass--relative-motion basis/representation as
the CM basis/representation.

The IP basis consists of products of eigenstates of the Hamiltonians
$H_1$ and $H_2$. In the 2D model the eigenstates of $H_i$ and
$l_z^{(i)}$ ($i = 1,2$) are the  Fock-Darwin
states $\Phi_{n_i,m_i}(\mathbf{r}_i)$ \cite{FD}.
 In the 3D model the elements of the IP basis are
$\Phi_{n_1,m_1}(\rho_1,\varphi_1)\, \phi_{n_{z1}}(z_1)\,
\Phi_{n_2,m_2}(\rho_2,\varphi_2)\, \phi_{n_{z2}}(z_2)$
(see Appendix~\ref{sec:IP-basis}).

In the IP representation the quantum number $M=m_1+m_2$ is a $z-$projection
of the total angular momentum.
Since $M$ is a good quantum number, the orbital wave function
$\psi(\mathbf{r}_1,\mathbf{r}_2)$ is expanded in the IP basis with
elements that are subject to the condition $m_1 + m_2 = M$. In the
2D model we have
\begin{equation}
\psi(\mathbf{r}_1,\mathbf{r}_2) = \sum_{n_1,n_2} \sum_{m_2=0}^M
a_{n_1,n_2,m_2}^{(M)} \Phi_{n_1,M-m_2}(\mathbf{r}_1)
\Phi_{n_2,m_2}(\mathbf{r}_2). \label{exp-ipb2d}
\end{equation}

The coefficients $a_{n_1,n_2,m_2}^{(M)}$, as well as the
corresponding eigenenergies, can be determined by diagonalizing
the Hamiltonian matrix $H$ in the IP (2D) basis. The interaction
matrix elements $\langle n_1, m_1, n_2, m_2|V_C|n_1^\prime,
m_1^\prime, n_2^\prime, m_2^\prime\rangle$ can be evaluated using,
for example, the expressions from Sec. 2.5.1. in
Ref.~\cite{chakraborty}. Of course, in numerical calculations the
basis must be restricted to a finite set ($n_1, n_2 =
0,\ldots,n_{\max}$) whose size is determined by the required accuracy
of results.
In the 3D case the expansion goes over the IP basis elements, that
includes the additional summation over the indices $n_{z1}$ and
$n_{z2}$. Since the Hamiltonian $H$ commutes
with the parity operator the orbital functions have a definite
parity.

For our analysis, it is convenient to use
symmetric and antisymmetric counterparts of the IP basis in the
orbital space. In the 2D case the corresponding basis elements are
the (anti)symmetrized products of the Fock-Darwin states, which we
will denote by $\{\Phi_{n_1,m_1}(\mathbf{r}_1),
\Phi_{n_2,m_2}(\mathbf{r}_2)\}_\pm$ (see Eq.~(\ref{ipb2d-sym}) in
Appendix~\ref{sec:IP-basis}). Symmetric ($+$) and
antisymmetric ($-$) eigenfunctions of the Hamiltonian $H$
are expanded as
\begin{eqnarray}
\psi^{(\pm)}(\mathbf{r}_1,\mathbf{r}_2) &=& \sum_{n_1,n_2}
\sum_{m_2=0}^{[M/2]} C_{n_1,n_2,m_2}^{(M,\pm)} \times \nonumber
\\
&& \{\Phi_{n_1,M-m_2}(\mathbf{r}_1),
\Phi_{n_2,m_2}(\mathbf{r}_2)\}_\pm, \label{exp2d-sym}
\end{eqnarray}
where $[M/2]$ is the integer part of $M/2$. This expression can be
readily generalized for the 3D model. In this case we use the
(anti)symmetrized products of the functions
$\Phi_{n_1,m_1}(\rho_1,\varphi_1)\, \phi_{n_{z1}}(z_1)$ and
$\Phi_{n_2,m_2}(\rho_2,\varphi_2)\, \phi_{n_{z2}}(z_2)$, and
perform the additional summation in Eq.~(\ref{exp2d-sym}) over the
indices $n_{z1}$ and $n_{z2}$.

The CM basis consists of the products of eigenstates of the Hamiltonians
$H_\mathrm{c.m.}$ and $H^{(0)}_\mathrm{rel}$.
These eigenfunctions are the Fock-Darwin states,
but with the values ${\cal M} = 2m^*$ and $\mu =
m^*/2$, respectively, instead of the effective electron mass $m^*$
(see Appendix~\ref{sec:CM-basis}).

The expansion of the orbital function $\psi(\mathbf{r}_1,
\mathbf{r}_2)$ in the CM basis goes only over the relative motion
quantum number(s) $n$ (and $n_z$ in the 3D model). This is a
consequence of the fact that the CM motion is fully separable in
cylindrical coordinates, and the quantum numbers $n_{c.m.}$,
$m_{c.m.}$ (and $n_z^{c.m.}$), and $m = M -
m_{c.m.}$, are good quantum numbers. Thus, in this
representation the eigenfunctions of the Hamiltonian $H$ are
factorized as
\begin{equation}
\psi(\mathbf{r}_1,\mathbf{r}_2) =
\psi_\mathrm{c.m.}(\mathbf{R})\,\psi_\mathrm{rel}(\mathbf{r}_{12}).
\label{psiorb}
\end{equation}
Here $\psi_\mathrm{c.m.} = \Phi_{n_{c.m.},
m_{c.m.}}^{(c.m.)}
(\rho_{c.m.},\varphi_{c.m.})\,
\phi_{n_z^{c.m.}}^{(c.m.)}(Z)$ are the eigenfunctions of
the Hamiltonian $H_\mathrm{c.m.}$, and
\begin{equation}
\psi_\mathrm{rel} = \sum_{n,n_z}\!\! b_{n,n_z}^{(m)}
\Phi_{n,m}^\mathrm{(rel)}(\rho_{12},\varphi_{12})\,
\phi_{n_z}^\mathrm{(rel)}(z_{12}) \label{psi_rm}
\end{equation}
are the eigenfunctions of the Hamiltonian $H_\mathrm{rel}$. In the
2D model the $\phi$-functions and the summation over $n_z$ is
removed. In contrast to the IP basis, the CM basis functions have
a definite exchange symmetry by construction (see
Appendix~\ref{sec:CM-basis}).

It seems likely that the CM representation is more convenient for
calculations. However, the physical essence of quantum electron
correlations is more transparent if the eigenfunctions are
expanded in the IP basis.

\subsection{Entanglement measures}

The entanglement of a pure state of a bipartite quantum system can
be quantified uniquely by the entropy of one of its subsystems.
Namely, if the global pure state $\Psi$ is entangled, i.e., it  cannot
be factorized into individual pure states of the subsystems, each
of the subsystems (1 and 2) is in a mixed state.
The degree of mixing can be determined by means
of the von Neumann entropy
$-\mathrm{Tr}(\rho_1\ln\rho_1) = -\mathrm{Tr}(\rho_2\ln\rho_2)$.
Here $\rho_1 =
\mathrm{Tr}_2|\Psi\rangle\langle\Psi|$ and $\rho_2 =
\mathrm{Tr}_1|\Psi\rangle\langle\Psi|$ are the reduced density
matrices, describing the mixed states of the subsystems 1 and 2,
respectively. Alternatively, one can calculate the linear entropy
${\cal E} = 1 - \mathrm{Tr}\rho_1^2 = 1 - \mathrm{Tr}\rho_2^2$
that can be obtained from the von Neumann entropy by expanding
the logarithm of the reduced density matrix and retaining the
leading term. Both entropies meet the necessary requirements for
an entanglement measure. The linear entropy is, however, more convenient
to compute, and we will use this measure.

For the systems consisting of two identical fermions
(electrons) the measure
must be modified in order to exclude the contribution
related to the antisymmetric character of fermionic states.
The correlations due to the Pauli principle do
not contribute to the state's entanglement. For instance, a
two-fermion state of a Slater rank 1 (i.e., a state
represented by one Slater determinant) must be regarded as
non-entangled, and its measure has to be zero.
In order to satisfy this requirement, the entanglement measure
based on the linear entropy in the case of two identical fermions
has the form (see, for example, \cite{NV,ber,YPD})
\begin{equation}
{\cal E} = \ 1 - 2\,\mathrm{Tr}\rho_r^2. \label{entm}
\end{equation}
Here $\rho_r$ is the reduced single-particle density matrix
obtained by tracing the two-particle density matrix $\rho =
|\Psi\rangle\langle\Psi|$ (describing the pure state $\Psi$) over
one of the two particles.
As a result, the measure (\ref{entm})  transforms to
the following form for a factorized  wave function ~(\ref{psitot})
\begin{equation}
{\cal E} = 1 - 2\,\mathrm{Tr}[{\rho_r^\mathrm{(orb)}}^2]\,
\mathrm{Tr}[{\rho_r^\mathrm{(spin)}}^2], \label{ent_measure}
\end{equation}
where $\rho_r^\mathrm{(orb)}$ and $\rho_r^\mathrm{(spin)}$ are the
reduced single-particle density matrices in the orbital and spin
spaces, respectively. In principle, the trace of ${\rho_r^\mathrm{(orb)}}^2$
 has to be calculated in some single-particle basis.
It is convenient, however,  to calculate this trace
by means of the formula
\begin{eqnarray}
\mathrm{Tr}[{\rho_r^\mathrm{(orb)}}^2] \!\!&=&\!\! \int
d\mathbf{r}_1\, d\mathbf{r}_1^{\,\,\prime}\, d\mathbf{r}_2\,
d\mathbf{r}_2^{\,\,\prime}\, \psi(\mathbf{r}_1,\mathbf{r}_2)\,
\psi^*(\mathbf{r}_1^{\,\,\prime},\mathbf{r}_2) \nonumber
\\[-.5ex]
&&\qquad\quad \psi^*(\mathbf{r}_1,\mathbf{r}_2^{\,\,\prime})\,
\psi(\mathbf{r}_1^{\,\,\prime},\mathbf{r}_2^{\,\,\prime}).
\label{trorb}
\end{eqnarray}
The trace of ${\rho_r^\mathrm{(spin)}}^2$ in the two-electron spin
states with a definite symmetry $\chi_{S,M_S}$ has
values: $1/2$ if $M_S = 0$ (antiparallel spins of two electrons),
and $1$ if $M_S = \pm1$ (parallel spins), i.e.,
\begin{equation}
\mathrm{Tr}[{\rho_r^\mathrm{(spin)}}^2] = \hbox{$\frac{1}{2}$}(1 +
|M_S|). \label{trace_spin}
\end{equation}

\section{The lowest states -- individual particle
description}\label{sec:indiv_pd}

In this section, using the individual particle picture, we
consider the lowest states of two-electron QDs in the magnetic field,
that are characterized by different values of the quantum number
$M$. These states represent the ground state of the system in
different intervals of the magnetic field strength (see
singlet-triplet transitions in Sec.~\ref{sec:ground_st}). We study
here the behaviour of these states in the limit of $V_C \to 0$.

\subsection{The lowest energy levels and states}

At $V_C =0$, by ignoring temporary the spin term $H_S$, the
two-electron energy levels are defined by the sum of
single-electron energies $E_{n_1,m_1,n_{z1}} + E_{n_2,m_2,n_{z2}}$
(see Eqs.~(\ref{fd_levels}), (\ref{H0lev3D})). The related
(unsymmetrized) eigenstates are the products
$\Phi_{n_1,m_1}(\rho_1,\varphi_1)\, \phi_{n_{z1}}(z_1)\,
\Phi_{n_2,m_2}(\rho_2,\varphi_2)\, \phi_{n_{z2}}(z_2)$ (i.e.
the IP basis, see Appendix~\ref{sec:IP-basis}).
The lowest levels are characterized by $n_1 = n_2 = n_{z1} =
n_{z2} = 0$ and $m_1, m_2 \ge 0$, and  are defined as
\begin{eqnarray}
E^{(0)}_M &=& E_{0,m_1,0} + E_{0,m_2,0} \nonumber
\\
&=& \hbar\Omega(M + 2) - \hbar\omega_L M + \hbar\omega_z,
\label{en_deg}
\end{eqnarray}
where $M = m_1 + m_2$ is the quantum number of the $z$-projection
of the total angular momentum.
The levels (\ref{en_deg}) are $M+1$ times degenerate
(orbital degeneracy). Thus, all linear combinations of the IP
basis states with $n_1 = n_2 = n_{z1} = n_{z2} = 0$ and $m_1 + m_2
= M$, or their (anti)symmetrized counterparts, are different
eigenstates of the same level $E^{(0)}_M$. Here, since all
two-electron states must have a definite exchange symmetry, it is
more convenient to use linear combinations with the
(anti)symmetrized basis functions.

\begin{figure}
\vspace{-.1in}
\epsfxsize 3in \epsffile{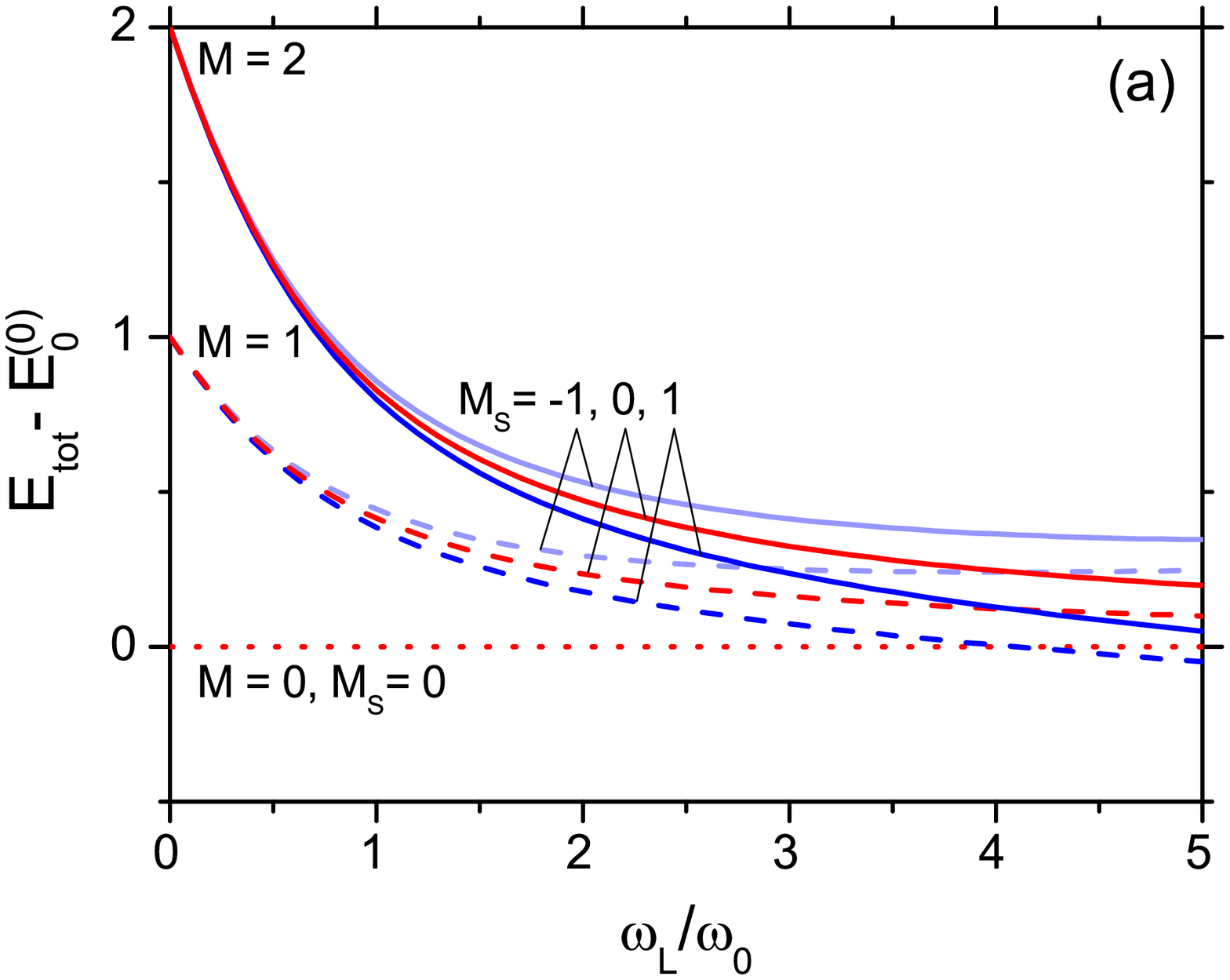}
\epsfxsize 3in \epsffile{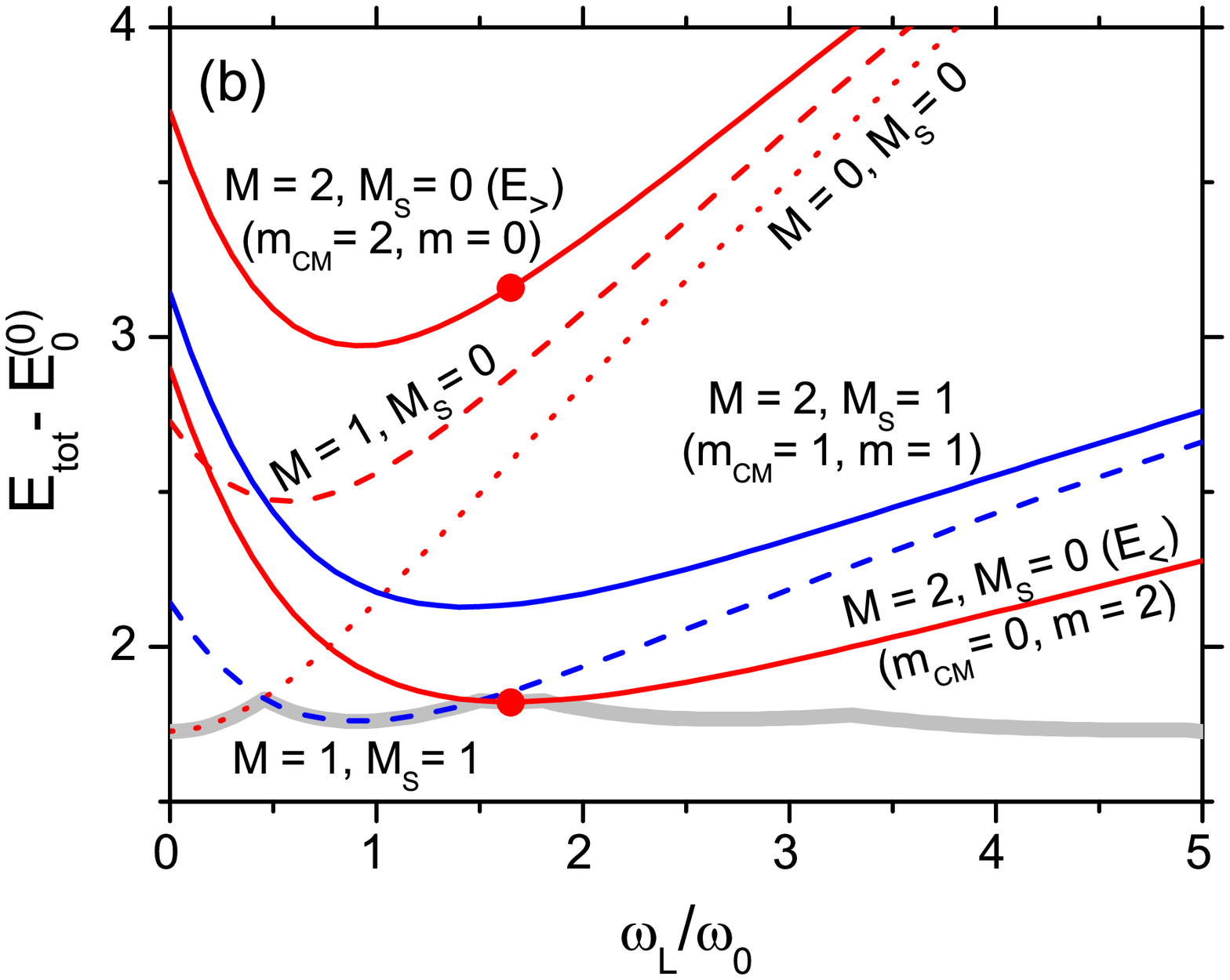}
\caption{(Color online) Lowest energy levels (in
$\hbar\omega_0$ units) of an axially symmetric two-electron QD
with $\omega_z \gg \omega_0$ for: (a) noninteracting electrons
($\lambda = 0$) and (b) $\lambda = 2$ (in $\hbar\omega_0\ell_0$
units). In order to get a better resolution the energy are defined with
respect to the
$E_0^{(0)}$ level. The Zeeman splitting is taken into
account: the effective Land\'e factor $g^* = -0.44$. Closed circles
mark the $M = 2$ and $M_S = 0$ levels at the field strength
$\omega_L/\omega_0 = 1.65$. At this value the lower level ($E_<$)
corresponds to the ground state energy (thick gray line).}
\label{fig:low_levels}
\end{figure}

For noninteracting electrons the
Zeeman splitting yields three different levels: $E_M^{(0)} + \Delta
E_{M_S}$, where $\Delta E_{M_S} = g^* \mu_B B M_S$ (see
Fig.~\ref{fig:low_levels}(a)). Obviously, for $g^* < 0$ ($g^* =
-0.44$ for GaAs) the lowest triplet level corresponds to $M_S =
1$, and we will further focus on the states with $M_S = S$.

Evidently, at $V_C \neq 0$, the orbital eigenfunctions are decomposed
in the IP basis in the form (\ref{exp-ipb2d}), or in the form (\ref{exp2d-sym}), where
the relation $m_1 + m_2 =M$ must be fulfilled for
different values of $n_1$, $n_2$, $n_{z1}$, $n_{z2}$.
The corresponding energy levels are
obtained by diagonalizing the Hamiltonian $H$ in the chosen
basis. One would expect, that  the lowest levels,
characterized by different values of $M$ ($E_M$), should converge
to the levels $E_M^{(0)}$ given by
Eq.~(\ref{en_deg}) in the limit $V_C \to 0$. The lowest levels (including the Zeeman
splitting) for $M = 0,1,2$ as functions of the magnetic field for a
typical axially symmetric two-electron QD are shown in
Fig.~\ref{fig:low_levels}(b).

\subsection{The limit of noninteracting electrons}
\label{sec:limit0_indpr}

As discussed above, at the limit $V_C \to 0$ it is natural to
expect that the lowest levels $E_M$ converge to the levels
$E_M^{(0)}$. The same must be true for the corresponding
eigenstates. This means that the expansion (\ref{exp2d-sym})
and its 3D counterpart are reduced to the sum of
contributions of the basis elements with $n_1 = n_2 = n_{z1} =
n_{z2} = 0$ and different values of $m_2 = M - m_1$ only.
In this case the $z$-component of the wave function is simply the
product $\phi_0(z_1) \phi_0(z_2)$ that is decoupled from
the lateral degrees of freedom. Consequently, quantum correlations
appear strictly due to coupling of functions (\ref{ipb2d-sym})
with $n_1 = n_2 = 0$. Hereafter (in this Section), for the sake of
simplicity, we will drop the $z$-component of the wave function
and, without loss of generality, consider the 2D description.

As a result, the expansion (\ref{exp2d-sym}) is reduced to the
sum
\begin{equation}
\psi^{(M,\pm)}(\mathbf{r}_1,\mathbf{r}_2) = \sum_{k=1}^{d_\pm}
c_{k}^{(M,\pm)} u^{(M,\pm)}_{k}(\mathbf{r}_1, \mathbf{r}_2),
\label{superposition}
\end{equation}
where the index $k = m_2 + 1$ counts the basis states
(\ref{ipb2d-sym}) of a given symmetry ($\pm$) with $n_1 = n_2 = 0$
and a given $M$, i.e.,
\begin{equation}
u^{(M,\pm)}_{k}(\mathbf{r}_1, \mathbf{r}_2) =
\{\Phi_{0,M-k+1}(\mathbf{r}_1), \Phi_{0,k-1}(\mathbf{r}_2)\}_\pm\,.
\label{ipb2d-sym0}
\end{equation}
Here
\begin{equation}
c_{k}^{(M,\pm)} = \lim_{V_C \to 0}C_{0,0,m_2}^{(M,\pm)}
\end{equation}
are the limiting values of the non-vanishing coefficients.
Occasionally, the notation $M$ in the superscript $(M,\pm)$ will
be dropped, when we deal with a fixed value of the quantum number
$M$. The explicit expressions for basis states (\ref{ipb2d-sym0})
for $M = 0,1,2,3$ are given in Table~\ref{table1}. The sum
(\ref{superposition}) has $d_\pm = [M/2] + 1$ terms, except the
antisymmetric case with an even $M$, when $d_- = [M/2]$ [because
$\{\Phi_{0,m_1} (\mathbf{r}_1, \Phi_{0,m_2} (\mathbf{r}_2)\}_{-}
= 0$ when $m_1 = m_2$]. In general, one can write $d_\pm = [M/2] +
1/2 + (\pm 1)^{M+1}/2$. These numbers can be treated as the
partial degeneracies of the level $E_M^{(0)}$, that take into
account only the states of a given exchange symmetry ($d_{+} +
d_{-} = M + 1$). The coefficients in the expansion
(\ref{superposition}) can be determined exactly by diagonalizing
the Hamiltonian $H$ (in the same limit) in the finite set ($k = 1,
\ldots, d_\pm$) of states (\ref{ipb2d-sym0}). In fact, in the
limit $V_C \to 0$ the full diagonalization procedure is reduced to
the first order degenerate perturbation approach. Below we study
the lowest states with $M = 0,1,2,3$ in the limit $V_C \to 0$.

\begin{table}
\caption{Basis states $u_{k}^{(M,\pm)}$ [Eq.~(\ref{ipb2d-sym0})]
for $M = 0,1,2,3$.} \label{table1}
\begin{center}
\begin{tabular}{cl}
\hline\hline
\\[-3ex]
$M$ & \qquad basis function
\\[.5ex]
\hline
\\[-2.5ex]
0 & $u_{1}^{(0,+)}(\mathbf{r}_1,\mathbf{r}_2) =
\Phi_{0,0}(\mathbf{r}_1) \Phi_{0,0}(\mathbf{r}_2)$
\\[.5ex]
\hline
\\[-2.5ex]
1 & $u_{1}^{(1,\pm)}(\mathbf{r}_1,\mathbf{r}_2) =
\frac{1}{\sqrt{2}} [\Phi_{0,1}(\mathbf{r}_1)
\Phi_{0,0}(\mathbf{r}_2) \pm \Phi_{0,0}(\mathbf{r}_1)
\Phi_{0,1}(\mathbf{r}_2)]$
\\[.5ex]
\hline
\\[-2.5ex]
  & $u_{1}^{(2,\pm)}(\mathbf{r}_1,\mathbf{r}_2) =
\frac{1}{\sqrt{2}}[\Phi_{0,2}(\mathbf{r}_1)
\Phi_{0,0}(\mathbf{r}_2) \pm \Phi_{0,0}(\mathbf{r}_1)
\Phi_{0,2}(\mathbf{r}_2)]$
\\[-1.5ex]
2 &
\\[-1.5ex]
  & $u_{2}^{(2,+)}(\mathbf{r}_1,\mathbf{r}_2) =
\Phi_{0,1}(\mathbf{r}_1) \Phi_{0,1}(\mathbf{r}_2)$
\\[.5ex]
\hline
\\[-2.5ex]
  & $u_{1}^{(3,\pm)}(\mathbf{r}_1,\mathbf{r}_2) =
\frac{1}{\sqrt{2}}[\Phi_{0,3}(\mathbf{r}_1)
\Phi_{0,0}(\mathbf{r}_2) \pm \Phi_{0,0}(\mathbf{r}_1)
\Phi_{0,3}(\mathbf{r}_2)]$
\\[-1.5ex]
3 &
\\[-1.5ex]
  & $u_{2}^{(3,\pm)}(\mathbf{r}_1,\mathbf{r}_2) =
\frac{1}{\sqrt{2}}[\Phi_{0,2}(\mathbf{r}_1)
\Phi_{0,1}(\mathbf{r}_2) \pm \Phi_{0,1}(\mathbf{r}_1)
\Phi_{0,2}(\mathbf{r}_2)]$
\\[1ex]
\hline\hline
\end{tabular}
\end{center}
\end{table}

\subsubsection{The lowest $M = 0$ state}

At $M=0$ the
magnetic quantum numbers of individual electrons are $m_1 = m_2 =
0$. The orbital degeneracy at $V_C=0$
 is 1 ($d_+ = 1$ and $d_- = 0$), and the only basis state
in the expansion (\ref{superposition}) is $u_1^{(+)}$ (see
Table~\ref{table1}). Therefore, the lowest $M = 0$ eigenstate of
$H$ in the limit $V_C \to 0$ is
\begin{equation}
\psi^{(+)}(\mathbf{r}_1,\mathbf{r}_2) =
u_1^{(+)}(\mathbf{r}_1,\mathbf{r}_2) \equiv
\Phi_{0,0}(\mathbf{r}_1) \Phi_{0,0}(\mathbf{r}_2). \label{psi0_M0}
\end{equation}
Due to the Pauli principle the related spin state is a singlet ($S
= M_S = 0$). Since it is antisymmetric, the total wave
function has the form of the Slater determinant, and the entanglement
of the lowest state with $M = 0$ must be zero.

\subsubsection{The lowest $M = 1$ states}

At $M = 1$, one has: ($m_1 = 1$, $m_2 = 0$) or ($m_1 = 0$, $m_2 =
1$). Thus, the orbital degeneracy is 2 ($d_+ = d_- = 1$).  We have
two basis states, $u_1^{(+)}$ and $u_1^{(-)}$ (see
Table~\ref{table1}), that may appear in the expansion
(\ref{superposition}). However, due to different symmetries these
states cannot create a superposition.  As a result, the lowest $M
= 1$ eigenstate is one of them, depending on the initially chosen
symmetry, i.e.,
\begin{eqnarray}
\psi^{(\pm)}(\mathbf{r}_1,\mathbf{r}_2) \!\!&=&\!\!
u_1^{(\pm)}(\mathbf{r}_1,\mathbf{r}_2) \label{psi0_M1}
\\
\!\!&\equiv&\!\! \frac{1}{\sqrt{2}} [\Phi_{0,1}(\mathbf{r}_1)
\Phi_{0,0}(\mathbf{r}_2) \pm \Phi_{0,0}(\mathbf{r}_1)
\Phi_{0,1}(\mathbf{r}_2)]. \nonumber
\end{eqnarray}
For $u_1^{(+)}$ the related spin state is the singlet, whereas for
$u_1^{(-)}$ we have the triplet state. Here we choose the
triplet state ($S = M_S = 1$) which is lower. Since this spin
state is the product of individual electron spin states with $m_s
= +1/2$, the total wave function of the lowest $M = 1$ state is a
Slater determinant and its entanglement is again zero.

\subsubsection{The lowest $M = 2$ states}

At $M = 2$ one has: ($m_1 = 2$, $m_2 = 0$), or ($m_1 = 1$, $m_2 =
1$), or ($m_1 = 0$, $m_2 = 2$). The orbital degeneracy is 3 ($d_+
= 2$, $d_- = 1$). As a result, we have three basis states: two
symmetric ($u_1^{(+)}$, $u_2^{(+)}$), and one antisymmetric
($u_1^{(-)}$) (see Table~\ref{table1}). The related spin states
are: singlet, singlet and triplet, respectively. At $V_C=0$ the
triplet state (due to the Zeeman splitting) produces a lower
energy (see Fig.~\ref{fig:low_levels}(a)). However, at $V_C\neq0$
and typical values of the effective Land\'e factor the lowest
energy level with $M = 2$ corresponds to the singlet state (see
Fig.~\ref{fig:low_levels}(b)). Due to this fact, at $M = 2$, we
will focus on the orbital symmetric states and study their
behaviour in the limit $V_C \to 0$.

\begin{figure}
\epsfxsize 3in \epsffile{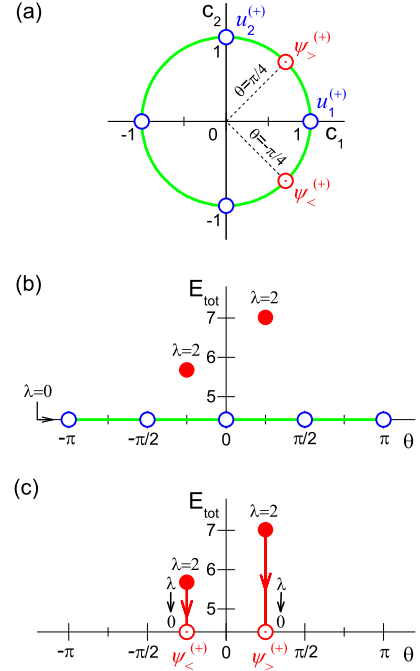}
\caption{(Color online) (a) Full range of (real) values of the
coefficients $c_1$ and $c_2$ in the superposition (\ref{psi_c1c2})
that represents the lowest symmetric orbital states $\psi^{(+)}$
with $M = 2$ at
$\lambda = 0$ (green circular line) and in the limit $\lambda \to 0$ (red
$\odot$ symbols). The values, marked by (blue) open circles,
correspond to the uncorrelated basis states $\pm
u_{1}^{(+)}$ and $\pm u_{2}^{(+)}$.
(b) Lowest $\psi^{(+)}$ states with $M = 2$ and the related
eigenenergies (in $\hbar\omega_0$ units) are shown in the
$\theta$-$E_\mathrm{tot}$ diagram, where $\theta =
\arctan(c_2/c_1)$, for $\lambda = 0$ (green line) and $\lambda =
2$ (red closed circles)(in $\hbar\omega_0\ell_0$ units). The full
$\theta$-domain ($\theta \in (-\pi,\pi)$), that corresponds to
$\lambda = 0$ (degenerate level $E_2^{(0)}$),  is reduced to two
discrete values ($\theta \approx \pm \pi/4$) as soon as
the interaction is switched on. The eigenenergies are calculated for
$\omega_L/\omega_0 = 1.65$ [at $\lambda = 2$ the state at
$\theta \approx -\pi/4$ is the ground state; see
Fig.~\ref{fig:low_levels}(b)]. (c) The $\theta$-$E_\mathrm{tot}$
diagram (the energy is in $\hbar\omega_0$ units)
showing the transition of the lowest (orbital symmetric)
states at $\lambda = 2$ to the states $\psi_\gtrless^{(+)}$ (see
Eq.~(\ref{psi0_M2})), obtained by reducing gradually the
interaction. The states remain entangled at any $\lambda \neq 0$
($\theta \approx \pm \pi/4$), even in the limit $\lambda \to 0$
(then $\theta = \pm \pi/4$).} \label{fig:c1c2}
\end{figure}

At $V_C=0$ the lowest
($M = 2$, $S = 0$) level is doubly degenerate ($d_+ = 2$).
As a result, the orbital wave function
may be any superposition of the states $u^{(+)}_{1}$ and
$u^{(+)}_{2}$, i.e.,
\begin{equation}
\psi^{(+)}(\mathbf{r}_1,\mathbf{r}_2) = c_1
u^{(+)}_{1}(\mathbf{r}_1,\mathbf{r}_2) + c_2
u^{(+)}_{2}(\mathbf{r}_1,\mathbf{r}_2)\,. \label{psi_c1c2}
\end{equation}
Here the coefficients $c_1$ and $c_2$ are arbitrary complex
numbers that are subject to  the condition $|c_1|^2 + |c_2|^2 = 1$ (green
circular line in Fig.~\ref{fig:c1c2}(a)). In this case it is
always possible to choose a set of eigenstates exhibiting zero
entanglement, e.g., $\psi_1^{(+)} = u_1^{(+)}$ and $\psi_2^{(+)} =
u_2^{(+)}$ (blue open circles) in
Fig.~\ref{fig:c1c2}(a)).

The nonzero interaction between the
electrons couples not only the functions $u_{1}^{(+)}$
and $u_{2}^{(+)}$, but also the symmetric functions
(\ref{ipb2d-sym}) with $m_1 + m_2 = M$ and $n_1, n_2 > 0$.
They will contribute to a given state
with specific values of the expansion coefficients in
(\ref{exp2d-sym}). For the lowest states,
however, the terms with $n_1 = n_2 = 0$ are dominant, even
at $V_C\neq 0$.
As a result,
the corresponding coefficients (here $C_{0,0,m_2}^{(2,+)}$, with $m_2 =
0,1$) practically determine the states (red closed circles in
Fig.~\ref{fig:c1c2}(b)).

By treating $V_C = \lambda/r_{12}$ as a small perturbation, the
coefficients $c_1$ and $c_2$ can be determined exactly in the
limit $\lambda \to 0$, using the first order degenerate
perturbation theory. One obtains the following set of linear
equations
\begin{eqnarray}
&& (\lambda V_{11} - \Delta E) c_1 + \lambda V_{12} c_2 =
0,\nonumber
\\[-1ex]
\label{c1c2eqs}
\\[-1ex]
&& \lambda V_{21} c_1 + (\lambda V_{22} - \Delta E) c_2 =
0,\nonumber
\end{eqnarray}
where $V_{ij} = \langle u_{i}^{(+)}| r_{12}^{-1}
|u_{j}^{(+)}\rangle$ and $\Delta E = E - E_2^{(0)}$. The
correction to the energy $\Delta E$ follows from the requirement
that the determinant of this system must be equal to zero (a secular
equation). As a result, one has
\begin{equation}
\Delta E_\gtrless = \frac{\lambda}{2}\left[V_{11} + V_{22} \pm
\sqrt{(V_{11}\!-\!V_{22})^2 + 4V_{12}V_{21}}\right].
\label{deltaE}
\end{equation}
Using one of Eqs.(\ref{c1c2eqs}) and the condition
$|c_1|^2 + |c_2|^2 = 1$, we obtain
\begin{eqnarray}
&& c_1 = \frac{V_{12}}{\sqrt{(V_{11} - \Delta E/\lambda)^2 +
V_{12}^2}}, \label{c1}
\\
&& c_2 = -\frac{V_{11} - \Delta E/\lambda}{\sqrt{(V_{11} - \Delta
E/\lambda)^2 + V_{12}^2}}. \label{c2}
\end{eqnarray}
The interaction matrix elements $V_{ij}$ can be numerically evaluated
 exactly (by means of the analytical expression for the
matrix elements $\langle m_1,m_2| r_{12}^{-1} |m_1^\prime,
m_2^\prime \rangle$ given in Ref.~\onlinecite{chakraborty}). For
$M = 2$ one has $V_{11} = V_{22} = 0.86165$, $V_{12} = V_{21} =
0.39166$. As a result, we obtain
\begin{equation}
\Delta E_\gtrless = \lambda(V_{11} \pm V_{12}) \label{deltaE_M2}
\end{equation}
and $c_1 = \pm c_2 = 1/\sqrt{2}$.
Thus, the related states are
\begin{eqnarray}
\psi_\gtrless^{(+)}(\mathbf{r}_1,\mathbf{r}_2) &\!\!=\!\!&
\frac{1}{\sqrt{2}}[u_{1}^{(+)}(\mathbf{r}_1,\mathbf{r}_2) \pm
u_{2}^{(+)}(\mathbf{r}_1,\mathbf{r}_2)] \nonumber
\\
&\!\!\equiv\!\!& \frac{1}{2}\, \Phi_{0,2}(\mathbf{r}_1)\,
\Phi_{0,0}(\mathbf{r}_2) \nonumber
\\
&\!\!\pm\!\!&\frac{1}{\sqrt{2}}\, \Phi_{0,1}(\mathbf{r}_1)\,
\Phi_{0,1}(\mathbf{r}_2) \label{psi0_M2}
\\
&\!\!+\!\!&\frac{1}{2}\, \Phi_{0,0}(\mathbf{r}_1)\,
\Phi_{0,2}(\mathbf{r}_2). \nonumber
\end{eqnarray}
In the diagrams in Fig.~\ref{fig:c1c2} these states are marked by
the symbol $\odot$. Note that the both solutions correspond to the
energy $E_2^{(0)}$ ($\Delta E_\gtrless \to 0$ in the limit
$\lambda \to 0$). However, if we start from the lowest ($M = 2$, $S =
0$) state of interacting electrons, the system  approaches the state $\psi_<^{(+)} =
(u_{1}^{(+)}\! - u_{2}^{(+)})/\sqrt{2}$ in the limit $\lambda \to 0$.

Thus, in the limit $V_C \to 0$ the lowest ($M = 2$, $S = 0$)
states will converge to the states (\ref{psi_c1c2}) with the specific
values of the coefficients $c_1$ and $c_2$ (the states
(\ref{psi0_M2})). In contrast, at $V_C = 0$ these coefficients
could be chosen to be arbitrary. The interaction removes the degeneracy
present in the noninteracting case, and leads the system (even in
the limit $V_C \to 0$) to an uniquely determined entangled state.
Indeed, it cannot be reduced to a single Slater determinant. This
is essentially the same mechanism, discussed in a qualitative
manner in Ref.~\onlinecite{MPD}.

\subsubsection{The lowest $M = 3$ states}

At $M = 3$ one has: ($m_1 = 3$, $m_2 = 0$), or ($m_1 = 2$, $m_2 =
1$), or ($m_1 = 1$, $m_2 = 2$), or ($m_1 = 0$, $m_2 = 3$). The
orbital degeneracy is 4 ($d_+ = d_- = 2$).  As a result, we have
four basis states, two symmetric and two antisymmetric:
$u_1^{(\pm)}$, $u_2^{(\pm)}$ (see Table~\ref{table1}). The orbital
functions $u_{i}^{(+)}$ and $u_{i}^{(-)}$ correspond to the
singlet and triplet spin states, respectively. For odd $M$
 the lowest energy level, for both 
noninteracting and interacting electrons, corresponds to the
triplet state with $M_S = 1$ (not shown in
Fig.~\ref{fig:low_levels}). Therefore, we consider
the  antisymmetric orbital states.

For noninteracting electrons the lowest level
($M = 3$, $S = M_S = 1$)
is doubly degenerate ($d_- = 2$). Therefore, the
orbital wave function may be any superposition of the states
$u_{1}^{(-)}$ and $u_{2}^{(-)}$ with arbitrary coefficients $c_1$
and $c_2$ that are subject to the condition $|c_1|^2 + |c_2|^2 = 1$. Again,
due to the degeneracy, it is possible to choose the
set of eigenstates, exhibiting a zero entanglement. For
interacting electrons, however, this state will converge
to the state $\psi^{(-)} = c_1 u_{1}^{(-)} + c_2 u_{2}^{(-)}$ with specific values of
the coefficients $c_1$ and $c_2$ in the limit $V_C \to 0$ .

Indeed, applying the first order degenerate perturbation theory,
by means of Eqs.(\ref{deltaE}), (\ref{c1}), (\ref{c2}), and the
corresponding values of the interaction matrix elements for $M =
3$ [$V_{11} = 0.56791$, $V_{22} = 0.45041$, $V_{12} = V_{21} =
\sqrt{3}\,(V_{11} - V_{22})/2 = 0.10176$], we obtain
\begin{equation}
\Delta E_\gtrless = \frac{\lambda}{2}[(V_{11} + V_{22}) \pm 2(V_{11} -
V_{22})]. \label{deltaE_M3}
\end{equation}
For $\Delta E_<$ we have $c_1 = 1/2$, $c_2 = -\sqrt{3}/2$;  and 
for $\Delta E_>$ we have $c_1 =\sqrt{3}/2$, $c_2 = 1/2$ . 
Thus, the related states are
\begin{eqnarray}
\psi_<^{(-)}(\mathbf{r}_1,\mathbf{r}_2) \!\!&=&\!\!
\frac{1}{2}[u_{1}^{(-)}(\mathbf{r}_1,\mathbf{r}_2) -
\sqrt{3}\,u_{2}^{(-)}(\mathbf{r}_1,\mathbf{r}_2)] \nonumber
\\
\!\!&\equiv&\!\! \frac{1}{2\sqrt{2}} \big[
\Phi_{0,3}(\mathbf{r}_1)\, \Phi_{0,0}(\mathbf{r}_2) \nonumber
\\
&& - \sqrt{3}\,\Phi_{0,2}(\mathbf{r}_1)\, \Phi_{0,1}(\mathbf{r}_2)
\label{psi0_M3I}
\\[1ex]
&& + \sqrt{3}\,\Phi_{0,1}(\mathbf{r}_1)\, \Phi_{0,2}(\mathbf{r}_2)
\nonumber
\\[1ex]
&& - \Phi_{0,0}(\mathbf{r}_1)\, \Phi_{0,3}(\mathbf{r}_2)\, \big],
\nonumber
\end{eqnarray}
\begin{eqnarray}
\psi_>^{(-)}(\mathbf{r}_1,\mathbf{r}_2) \!\!&=&\!\!
\frac{1}{2}[\sqrt{3}\,u_{1}^{(-)}(\mathbf{r}_1,\mathbf{r}_2) +
u_{2}^{(-)}(\mathbf{r}_1,\mathbf{r}_2)] \nonumber
\\
\!\!&\equiv&\!\! \frac{1}{2\sqrt{2}} \big[
\sqrt{3}\,\Phi_{0,3}(\mathbf{r}_1)\, \Phi_{0,0}(\mathbf{r}_2)
\nonumber
\\
&& + \Phi_{0,2}(\mathbf{r}_1)\, \Phi_{0,1}(\mathbf{r}_2)
\label{psi0_M3II}
\\[1ex]
&& - \Phi_{0,1}(\mathbf{r}_1)\, \Phi_{0,2}(\mathbf{r}_2) \nonumber
\\[1ex]
&& - \sqrt{3}\,\Phi_{0,0}(\mathbf{r}_1)\,
\Phi_{0,3}(\mathbf{r}_2)\, \big]. \nonumber
\end{eqnarray}
Note that the both solutions correspond to the energy $E_3^{(0)}$.
However, if we start from the lowest ($M = 3$, $S = M_S = 1$)
state of interacting electrons, the system in the limit $\lambda
\to 0$ approaches the state $\psi_<^{(-)} = (u_{1}^{(-)}\! -
\sqrt{3}\,u_{2}^{(-)})/2$. As in the previous case, the
lowest $M = 3$ state is entangled because it cannot be reduced to
a single Slater determinant.

\subsection{Entanglement of the lowest states in the limit $V_C \to 0$}

In the general case (within the 2D model) the
lowest eigenstates with $M = 0, 1, 2, 3$ of the Hamiltonian $H$ in
the IP basis can be written
in the form
\begin{equation}
\psi^{(\pm)}(\mathbf{r}_1,\mathbf{r}_2) = \sum_{m_2=0}^M
a_{m_2}^{(\pm)} \Phi_{0,M-m_2}(\mathbf{r}_1)
\Phi_{0,m_2}(\mathbf{r}_2), \label{exp-ipb2d0}
\end{equation}
where the coefficients $a_{m_2}^{(\pm)}$ can be expressed in terms
of the coefficients $c_k^{(\pm)}$ in the expansion
(\ref{superposition}). The reduced density matrix of the
state (\ref{exp-ipb2d}), calculated in the single-particle
Fock-Darwin basis, will be a sum consisting of $M+1$ terms. For
example,
\begin{equation}
\rho_1^\mathrm{(orb)} = \mathrm{Tr}_2\,|\psi\rangle\langle\psi| =
\sum_{m_2=0}^M{\big._2\!}\langle
m_2|\psi\rangle\langle\psi|m_2\rangle{\big._2},
\end{equation}
where $_2\langle m_2|\psi\rangle$ denotes the (partial) scalar
product between the Fock-Darwin state $\Phi_{0,m_2}(\mathbf{r}_2)$
and the orbital state $\psi(\mathbf{r}_1,\mathbf{r}_2)$. For the
state (\ref{exp-ipb2d0}) one has $_2\langle
m_2|\psi^{(\pm)}\rangle = a_{m_2}^{(\pm)} |M\!-\!m_2\rangle_1 =
a_{m_2}^{(\pm)} |m_1\rangle_1$, where $m_1 = M\!-\!m_2$. Here the
Dirac ket $|m_1\rangle_1$ corresponds to the Fock-Darwin state
$\Phi_{0,m_1}(\mathbf{r}_1)$. Thus, we have
\begin{eqnarray}
\rho_1^\mathrm{(orb)} &\!\!=\!\!& \sum_{m_2=0}^M
\big|a_{m_2}^{(\pm)}\big|^2 |M\!-\!m_2\rangle{\big._1}
{\big._1}\!\langle M\!-\!m_2| \nonumber
\\
&\!\!=\!\!& \sum_{m_1=0}^M \big|a_{M-m_1}^{(\pm)}\big|^2
|m_1\rangle{\big._1} {\big._1}\!\langle m_1|.
\end{eqnarray}
In fact, the reduced density matrix is diagonal in the
Fock-Darwin basis. One obtains readily  its square
\begin{equation}
{\rho_1^\mathrm{(orb)}}^2 = \sum_{m_1=0}^M \big|a_{M-m_1}^{(\pm)}
\big|^4 |m_1\rangle{\big._1} {\big._1}\!\langle m_1|
\end{equation}
and the trace
\begin{eqnarray}
\mathrm{Tr}[{\rho_r^\mathrm{(orb)}}^2] &\!\!=\!\!& \sum_{m_1=0}^M
\big|a_{M-m_1}^{(\pm)}\big|^4 = \sum_{m_2=0}^M
\big|a_{m_2}^{(\pm)}\big|^4. \label{trace_orb0}
\end{eqnarray}
Finally, the entanglement measure $\cal E$ can be obtained by
means of Eqs.~(\ref{ent_measure}), (\ref{trace_spin}),
(\ref{trace_orb0}).

The results for $M = 0,1,2,3$ are shown in Table~\ref{table2}. One
can see that, in agreement with the analysis from the previous
subsection, in the case of vanishing electron-electron interaction
($V_C \to 0$) the lowest states with $M = 0,1$ are not entangled
(${\cal E} = 0$), while those with $M = 2,3$ are entangled (${\cal
E} > 0$).

We would like to point out that the coefficients $a_{m_2}^{(\pm)}$
determine solely the trace (\ref{trace_orb0}), i.e., the
entanglement measure $\cal E$. In appears that the decomposition
(\ref{exp-ipb2d0}) in terms of the Fock-Darwin states is similar
to the Schmidt decomposition for a helium atom (see, e.g.,
Refs.~\onlinecite{h1,h2}).

\begin{table}
\caption{Entanglement measure based on the linear entropy
$\cal E$ and the corresponding orbital and spin factors (traces)
for the lowest states with $M = 0,1,2,3$ in the limit of
noninteracting electrons.}\label{table2}
\begin{center}
\begin{tabular}{ccccc}
\hline\hline
\\[-2.5ex]
$M$ & $S(M_S)$ & $\mathrm{Tr}[{\rho_r^\mathrm{(orb)}}^2]$ &
$\mathrm{Tr}[{\rho_r^\mathrm{(spin)}}^2]$ & $\cal E$
\\[.5ex]
\hline
\\[-2.5ex]
0 & 0 & $1$ & $1/2$ & $0$
\\[-.5ex]
1 & 1 & $1/2$ & $1$ & $0$
\\[-.5ex]
2 & 0 & $3/8$ & $1/2$ & $5/8$
\\[-.5ex]
3 & 1 & $5/16$ & $1$ & $3/8$
\\[.5ex]
\hline\hline
\end{tabular}
\end{center}
\end{table}

\section{The lowest states in the CM representation}
\label{sec:CM-rep}

\subsection{Classification of the lowest states}

In the CM basis  the determination of the wave function
$\psi(\mathbf{r}_1, \mathbf{r}_2)$ is reduced to the calculation
of $\psi_\mathrm{rel} (\mathbf{r}_{12})$ in the form of the expansion
(\ref{psi_rm}). In this representation each state is characterized
by the CM quantum numbers: $n_{c.m.}$, $m_{c.m.}$ and
(in the 3D model) $n_z^{c.m.}$, as well as by the magnetic
quantum number for the relative motion $m$.
The total quantum numbers $M$, $S$ and $M_S$ (and the related
exchange symmetry and parity) are good quantum numbers as well as
in the IP representation.

The quantum numbers $m_{c.m.}$, $m$ and $M$, however,
are not independent ($M = m_{c.m.} + m$), and for labeling
the states it is sufficient to use two of them. Moreover, in the
lowest states also the values of the spin quantum numbers $S$ and
$M_S$ depend on $m$. Namely, for these states the parity of the
wave function $\psi_\mathrm{rel}$ is $(-1)^m$. This rule holds
even in the 3D model, since the leading term in the expansion
(\ref{psi_rm}) is $\Phi^\mathrm{(rel)}_{0,m}\,
\phi^\mathrm{(rel)}_0$ and, thus, all $n_z$ must be even.  As a
result, due to the Pauli principle, the total spin is determined
by the expression $S = \frac{1}{2}[1 - (-1)^m]$.

The quantum number $M_S$, on the other hand, is not directly
determined by the parity. For even $m$ we have $S = 0$
and, thus, $M_S= 0$. If $m$ is odd, we have $S = 1$, and
$M_S$ can be $-1$, $0$, or $1$.
At nonzero magnetic field the Zeeman splitting
($g^* < 0)$ will
lower the energy of the $M_S = 1$ component of the triplet states,
while leaving the singlet states unchanged. Therefore, the lowest
states are characterized by
\begin{equation}
M_S = S = \frac{1 - (-1)^m}{2}. \label{M_S}
\end{equation}
Consequently, the lowest states (by considering various values of
$M$) can be uniquely labeled by the pair of quantum numbers
$(M,m)$ or by the pair $(m_{c.m.},m)$ (here $n_{c.m.} =
n_z^{c.m.} = 0$).

At $V_C=0$ the quantum
numbers $m_{c.m.}$ and $m$ play symmetric roles [analogously
to the quantum numbers $m_1$ and $m_2$ in Eq.~(\ref{en_deg})].
As a result,
one obtains that the energy levels $E_M^{(0)}$ (here constructed
as $E_{c.m.} + E_\mathrm{rel}^{(0)}$) are $M + 1$ times
degenerate. However, at $V_C \neq 0$ the spectra of the Hamiltonians
$H_{c.m.}$ and $H_\mathrm{rel}$ are different, and the
degeneracy is removed. In this case the lowest states correspond
to $m = M$ (i.e. $m_{c.m.} = 0$), as one can see from the
example shown in Fig.~\ref{fig:low_levels}(b). This feature can be
explained by the fact that the energy contribution due to the
Coulomb interaction $\Delta E^\textsc{(c)}$ in $E_\mathrm{rel}$
($E_\mathrm{rel} = E_\mathrm{rel}^{(0)} + \Delta E^\textsc{(c)}$)
decreases if $m$ increases. In other words, a rotational
motion of electrons attenuates the interaction.
Thus, for a given $M$ the total
energy $E_{M,m} = E_M^{(0)} + \Delta E^\textsc{(c)}_m$ is minimal
at $m = M$.

\subsection{Calculation of the measure $\cal E$}

In a general case it is convenient to calculate the entanglement
measure (\ref{ent_measure}) by evaluating the integral
(\ref{trorb}) and the expression (\ref{trace_spin}).
Combining the later with Eq.~(\ref{M_S}), we obtain for the lowest
states the expression for the spin contribution
\begin{equation}
\mathrm{Tr}[{\rho_r^\mathrm{(spin)}}^2] = \frac{3 - (-1)^m}{4}.
\label{trspin_lowst}
\end{equation}

The orbital contribution (\ref{trorb}) of the
lowest CM eigenstate ($n_{c.m.} = m_{c.m.} =
n_z^{c.m.} = 0$), applying the factorization (\ref{psiorb})
and the expansion (\ref{psi_rm}), takes the form
\begin{eqnarray}
\mathrm{Tr}[{\rho_r^\mathrm{(orb)}}^2] \!\!\!&=&\!\!\!
\sum_{n_1=0}^{n_{\max}} \sum_{n_2=0}^{n_{\max}}
\sum_{n_3=0}^{n_{\max}} \sum_{n_4=0}^{n_{\max}}
\sum_{n_{z_1\!}=0}^{n_z^{\max}} \sum_{n_{z_2}\!=0}^{n_z^{\max}}
\sum_{n_{z_3}\!=0}^{n_z^{\max}} \sum_{n_{z_4}\!=0}^{n_z^{\max}}
\nonumber
\\
&&\qquad b_{n_1,n_{z_1}}^{(m)} b_{n_2,n_{z_2}}^{(m)}
b_{n_3,n_{z_3}}^{(m)} b_{n_4,n_{z_4}}^{(m)} \times \nonumber
\\
&&\qquad\qquad I(n_1,n_2,n_3,n_4;m)\times \nonumber
\\
&&\qquad\qquad J(n_{z_1},n_{z_2},n_{z_3},n_{z_4}), \label{trace3d}
\end{eqnarray}
where $I$ and $J$ are multiple integrals of the products of
four functions $\Phi_{n,m}^{(c.m.)}$ and
$\phi_{n_z}^{(c.m.)}$ with different values of the indices
$n$ and $n_z$, respectively. Their explicit forms are given in
Appendix \ref{sec:IJ-integrals}.

In the 2D model the trace (\ref{trace3d}) is reduced to the form
with four sums via $n_i$ (i = 1,2,3,4). Formally it follows if we
put $n_z^{\max} = 0$ and drop the $J$ integrals (because
$J(0,0,0,0) = 1$).

The values of $I$ and $J$ integrals can be determined
analytically. In particular, for $n_1 = n_2 = n_3 = n_4 = 0$ one
has
\begin{equation}
I(0,0,0,0;m) = \frac{(2|m|)!}{(2^{|m|} |m|!)^2}. \label{I0}
\end{equation}
Generally, the $I$ and $J$ integrals do not depend on the QD
parameters (e.g. $\Omega$ and $\omega_z$, see Appendix
\ref{sec:IJ-integrals}). Thus the variation of the entanglement
 ${\cal E}(\Omega)$ is brought
about by the expansion coefficients in (\ref{psi_rm})
that depend on the interaction and the magnetic field strengths.
Some properties of the $I$ and $J$ integrals are given in Appendix
\ref{sec:IJ-integrals}.

\subsection{The dependance of the measure ${\cal E}$ on
the electron-electron interaction strength}

\begin{figure}
\epsfxsize 3.2in \epsffile{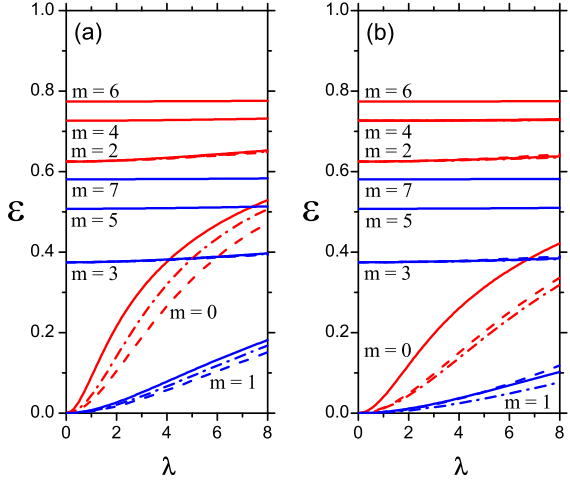}
\caption{(Color online) Entanglement measure $\cal E$ of the
lowest states with different $m$
as functions of the  of electron-electron interaction strength
(the parameter $\lambda$ in $\hbar\omega_0 \ell_0$ units): 
for (a) $\omega_L = 0$ and (b) 
$\omega_L/\omega_0 = 2.5$. The dashed, dash-dotted and solid lines
correspond to the (3D) QDs with $\omega_z/\omega_0 = 2$ and
$\omega_z/\omega_0 = 5$ and to the 2D model of QD, respectively.
The red and the blue values correspond to the symmetric and 
antisymmetric orbital states (i.e., even and odd
$\psi_\mathrm{rel}(\mathbf{r}_{12})$ functions), respectively.}
\label{fig:ent-rw}
\end{figure}

The measure $\cal E$ as a function of $\lambda$ for
the lowest states is  obtained by calculating the trace (\ref{trace3d}) with $n_{\max} =
4$ and $n_z^{\max} = 0$ (the 2D model) and $n_z^{\max} = 4$ (the 3D
model) (see Fig.~\ref{fig:ent-rw}). This basis size is found to be sufficient for the lowest
states.

In the limit $\lambda \to 0$ the measure $\cal E$ vanishes for $m
= 0$ and $m = 1$ and converges to non-zero values for $m \ge 2$,
independently of the magnetic field strength. This result is in
agreement with the results of Sec.~\ref{sec:indiv_pd} (see
Table~\ref{table2}). With the increase of the interaction strength
$\lambda$ the measure $\cal E$  increases, since the interaction
introduces additional correlations. However, for the states with
larger values of $m$ this change is slow. It appears that for typical QDs
($\lambda \sim 2\!\times\!\hbar\omega_0 l_0$) the measure $\cal E$
for $m \ge 2$ can be approximated by its value obtained at the
noninteracting case.

The entanglement of
the lowest states depends on the ratio $\omega_z/\omega_0$.
This dependence is very weak for the states with
large $m$ (see Fig.~\ref{fig:ent-rw}).
This means that for typical QDs the 2D model is good
enough if we calculate the entanglement of the lowest states with
$m \ge 2$. On the other hand, for the lowest states with $m = 0$ and $m = 1$
this model may be insufficient.

At $B = 0$ the entanglement decreases if the ratio
$\omega_z/\omega_0$ decreases from $\infty$ (the 2D model) to $1$
(the spherically symmetric 3D model); see Fig.~\ref{fig:ent-rw}(a) and
Fig.~\ref{fig:ent-oml} at $\omega_L/\omega_0 = 0$. This effect can
be explained by introducing the  effective charge
$\lambda_\mathrm{eff}$ \cite{sc,effch}. As pointed out in
Ref.~\onlinecite{sc}, in a 3D dot the electrons can avoid each
other more effectively than in the corresponding 2D one. As a
result, the Coulomb interaction has a smaller effect on the 3D
spectrum than on the 2D one. The ratio
$\lambda_\mathrm{eff}/\lambda$ as a function of $\Omega/\omega_z$
for different $m$ is shown in Fig.~2 in Ref.~\onlinecite{effch}.
The maximal repulsion at $\Omega/\omega_z = 0$ corresponds with
$\omega_z \to \infty$ to the 2D case and decreases monotonically
as $\Omega/\omega_z$ increases. Thus, a reduction of the ratio
$\omega_z/\omega_0$ has an effect analogous to the reduction of
the electron-electron interaction.

If $B \neq 0$, the entanglement measure $\cal E$ is not
necessarily a monotonic function of the ratio $\omega_z/\omega_0$.
In contrast to the zero-magnetic-field case, for example, at
$\omega_L/\omega_0 = 2.5$ [see
Fig.~\ref{fig:ent-rw}(b)] the lowest states are more entangled
for the ratio $\omega_z/\omega_0 = 2$ (dashed lines) in comparison
with
the ratio $\omega_z/\omega_0 =5$ (dash-dotted lines).
This behavior can be understood by analyzing the dependence of
the entanglement of the lowest states on the magnetic field (see the
next subsection).
Regarding the latest example, Fig.~\ref{fig:ent-oml} shows how the
values of $\cal E$ for $\omega_z/\omega_0 = 2$ and
$= 5$ exchange the order between
$\omega_L/\omega_0 = 0$ and $= 2.5$ 
(the case $m= 0$ is shown).

\subsection{The dependence of the measure ${\cal E}$ on the magnetic field strength}

\begin{figure}
\vspace{.04in} \epsfxsize 2.8in \epsffile{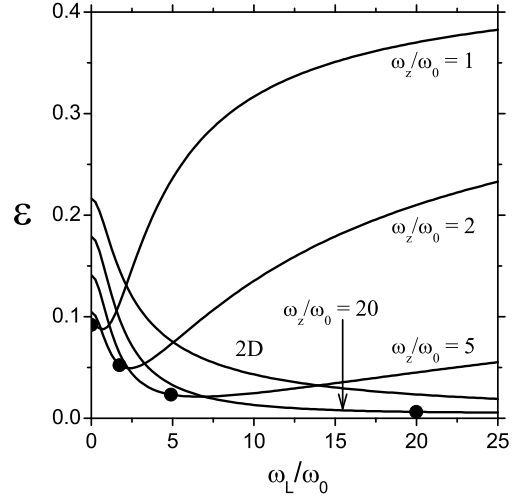}
\caption{Entanglement of the lowest state with $m = 0$ of axially
symmetric two-electron QDs with $\lambda = 2$ (in $\hbar\omega_0
\ell_0$ units) and different ratios $\omega_z/\omega_0$ as a
function of the parameter $\omega_L/\omega_0$. The closed circles
denote the values of $\omega_L/\omega_0$ when the dots with the
given ratios $\omega_z/\omega_0$ become spherically symmetric.}
\label{fig:ent-oml}
\end{figure}

For extremely thin QDs (described with the aid of the 2D model),
the entanglement of the lowest states decreases monotonically by
increasing the magnetic field strength (see
Fig.~\ref{fig:ent-oml}). Evidently, the effective confinement
($\Omega$) increases with the magnetic field, and the contribution
of the constant electron-electron interaction becomes weaker.
Formally, if we introduce the characteristic length of the
effective confinement $\ell_\Omega = \sqrt{\hbar/m^*\Omega}$, the
parameter $\lambda_\Omega = \ell_\Omega/a^*$ ($R_W$ at $B = 0$)
determines the relative strength of the Coulomb interaction at a given
effective confinement \cite{PSN}. This parameter decreases  with
an increase of the magnetic field strength. It tends to zero at
$B \to \infty$, since the Coulomb interaction becomes negligible
compared to the effective confinement.

 In 3D cases, however, with an increase of the magnetic
field the entanglement decreases
 until the parameter $\omega_L$ reaches the value
$\omega_L^\mathrm{sph} = (\omega_z^2 - \omega_0^2)^{1/2}$. At this
value the QD owns a spherically symmetry ($\Omega/\omega_z = 1$).
After this value the measure starts to increase with an increase
of the magnetic field (see Fig.~\ref{fig:ent-oml}). This behavior
can be explained by the influence of the magnetic field on the
effective strength $\lambda_\Omega$. It is twofold -- the magnetic
field, besides the effective confinement, affects the effective
charge, too. In fact, the entanglement indicates a
geometrical crossover in two-electron QDs. In particular, the
lateral electron density distribution transforms to the vertical
one for a state with a given value of the magnetic quantum
number $m$. A detailed analysis of this phenomenon
for $m=0$ state is presented in
Refs.~\onlinecite{NSPC,NS13}.

Note that the effect of the magnetic field on the measure $\cal E$ for
states with large quantum number $m$ is less pronounced than for
the $m = 0$ state (see Fig.~\ref{fig:entgr2d}(b) for the 2D case and in
Fig.~\ref{fig:entgr3d} for 3D case).

\subsection{The limit of noninteracting electrons}

At $V_C=0$ the lowest
eigenstates of the Hamiltonian $H_\mathrm{rel}$ converge to the
lowest eigenstates of the Hamiltonian $H_\mathrm{rel}^{(0)}$, and
the orbital wave functions (\ref{psiorb}) take the form
\begin{equation}
\psi = \Phi^{(c.m.)}_{0,0} \phi^{(c.m.)}_0
\Phi^\mathrm{(rel)}_{0,m} \phi^\mathrm{(rel)}_0. \label{low_st0}
\end{equation}
In this limit all coefficients in the expansion (\ref{psi_rm})
tend to zero, except the first one ($b_{0,0}^{(m)}$)
that tends to one. As a result,  Eq.~(\ref{trace3d}) is reduced to a
single term
\begin{equation}
\mathrm{Tr}[{\rho_r^\mathrm{(orb)}}^2] = I(0,0,0,0;m)
\label{trorb_lowst}
\end{equation}
(here we used $J(0,0,0,0) = 1$).  By means of
Eqs. (\ref{ent_measure}), (\ref{I0}),
(\ref{trorb_lowst}), (\ref{trspin_lowst}), we obtain the
result
\begin{equation}
{\cal E} = 1 - \frac{3 \!-\! (-1)^m}{2}\, \frac{(2m)!}{(2^m\,
m!)^2}\,, \label{entm_nint}
\end{equation}
that exactly reproduces the results given in 
Table~\ref{table2} (here $M = m$).  The formula (\ref{entm_nint})
is valid for the 2D and the 3D models of QDs.

\begin{figure}
\epsfxsize 3in \epsffile{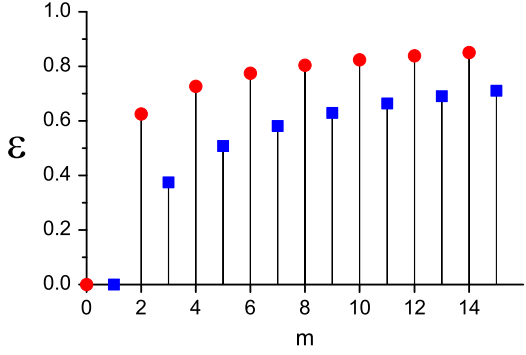}
\caption{(Color online) Measure of entanglement $\cal E$ of the lowest states
($n_{c.m.} = n_z^{c.m.} = m_{c.m.} = 0$, $n =
n_z = 0$) for various $m$
in the limit of noninteracting electrons.} \label{fig:entm_nint}
\end{figure}

The entanglement increases differently for even and odd values
of $m$  (see  Fig.~\ref{fig:entm_nint}).
This staggering is due to the spin contribution
(\ref{trspin_lowst}), whereas the orbital part is responsible for
the growth of $\cal E$ with $m$. Finally, since $I(0,0,0,0;m) \to
0$ when $m \to \infty$, both series converge to $1$, i.e.,
\begin{equation}
\lim_{m \to \infty} {\cal E} = 1.
\end{equation}

\section{The connection between representations}
\label{sec:CM-Rel-rep}

As already mentioned,  due to the separation (\ref{psiorb}),
the eigenstates of the Hamiltonian $H$ have a simpler form in the
CM representation than in the IP one. However, the simplicity of this
form holds back information on the inherent correlations of the
two-electron orbital eigenfunctions $\psi(\mathbf{r}_1,
\mathbf{r}_2)$. To illuminate this structure, one has to employ
the IP representation.

\begin{table*}
\caption{Individual particle and CM representations of the
orbital parts of lowest states (including those with
$m_{c.m.} > 0$) of axially symmetric two-electron QDs (in the
magnetic field) in the limit of noninteracting electrons ($V_C \to
0$) for $M = 0,1,2,3$, the related spin states ($S,M_S$), the
traces of squares of the corresponding reduced density matrices
and the values of the entanglement measure $\cal E$. The state,
that is the lowest for a given $M$, corresponds to $m_{c.m.}
= 0$ (shown on the top of each $M$-manifold) and $M_S = S$. The
related values of $\cal E$ (highlighted) coincide with those given
in Table~\ref{table2}.}\label{table3}
\begin{center}
\begin{tabular}{ccccccccc}
\hline\hline
\\[-2.5ex]
$M$ & orb. state & \hspace{.5in} ind. particle rep. \hspace{.5in}
& \hspace{.2in} CM rep. \hspace{.2in} &
$\mathrm{Tr}[{\rho_r^\mathrm{(orb)}}^2]$ & \;$S$\; & $M_S$ &
$\mathrm{Tr}[{\rho_r^\mathrm{(spin)}}^2]$ & \hspace{.05in} $\cal
E$ \hspace{.05in}
\\[.5ex]
\hline
\\[-2.5ex]
0& $\psi^{(0,+)}$ & $u_{1}^{(0,+)}(\mathbf{r}_1,\mathbf{r}_2)$ &
$\Phi^{(c.m.)}_{0,0}(\mathbf{R})
\Phi^\mathrm{(rel)}_{0,0}(\mathbf{r}_{12})$ & 1 & 0 & 0 & $1/2$ &
$\mathbf{0}$
\\[.5ex]
\hline
\\[-2.5ex]
& $\psi^{(1,-)}$ & $u_{1}^{(1,-)}(\mathbf{r}_1,\mathbf{r}_2)$ &
$\Phi^{(c.m.)}_{0,0}(\mathbf{R})
\Phi^\mathrm{(rel)}_{0,1}(\mathbf{r}_{12})$ & 1/2 & 1 & $0,\pm1$ &
$1/2,1$ & $1/2,\,\mathbf{0}$
\\[-1.5ex]
1 & & &
\\[-1.5ex]
& $\psi^{(1,+)}$ & $u_{1}^{(1,+)}(\mathbf{r}_1,\mathbf{r}_2)$ &
$\Phi^{(c.m.)}_{0,1}(\mathbf{R})
\Phi^\mathrm{(rel)}_{0,0}(\mathbf{r}_{12})$ & 1/2 & 0 & 0 & 1/2 &
1/2
\\[.5ex]
\hline
\\[-2.5ex]
& $\psi_{<}^{(2,+)}$ &
$\frac{1}{\sqrt{2}}[u_{1}^{(2,+)}(\mathbf{r}_1,\mathbf{r}_2) -
u_{2}^{(2,+)}(\mathbf{r}_1,\mathbf{r}_2)] $ &
$\Phi^{(c.m.)}_{0,0}(\mathbf{R})
\Phi^\mathrm{(rel)}_{0,2}(\mathbf{r}_{12})$ & 3/8 & 0 & 0 & 1/2 &
$\mathbf{5/8}$
\\[.5ex]
2 & $\psi^{(2,-)}$ & $u_{1}^{(2,-)}(\mathbf{r}_1,\mathbf{r}_2)$ &
$\Phi^{(c.m.)}_{0,1}(\mathbf{R})
\Phi^\mathrm{(rel)}_{0,1}(\mathbf{r}_{12})$ & 1/2 & 1 & $0,\pm1$ &
$1/2,1$ & $1/2,\,0$
\\[.5ex]
& $\psi_{>}^{(2,+)}$ &
$\frac{1}{\sqrt{2}}[u_{1}^{(2,+)}(\mathbf{r}_1,\mathbf{r}_2) +
u_{2}^{(2,+)}(\mathbf{r}_1,\mathbf{r}_2)] $ &
$\Phi^{(c.m.)}_{0,2}(\mathbf{R})
\Phi^\mathrm{(rel)}_{0,0}(\mathbf{r}_{12})$ & 3/8 & 0 & 0 & 1/2 &
5/8
\\[.5ex]
\hline
\\[-2.5ex]
& $\psi_{<}^{(3,-)}$ &
$\frac{1}{2}[u_{1}^{(3,-)}(\mathbf{r}_1,\mathbf{r}_2) -
\sqrt{3}\,u_{2}^{(3,-)}(\mathbf{r}_1,\mathbf{r}_2)]$ &
$\Phi^{(c.m.)}_{0,0}(\mathbf{R})
\Phi^\mathrm{(rel)}_{0,3}(\mathbf{r}_{12})$ & 5/16 & 1 & $0,\pm1$
& $1/2,1$ & $11/16,\,\mathbf{3/8}$
\\[.5ex]
& $\psi_{<}^{(3,+)}$ &
$\frac{1}{2}[\sqrt{3}\,u_{1}^{(3,+)}(\mathbf{r}_1,\mathbf{r}_2) -
u_{2}^{(3,+)}(\mathbf{r}_1,\mathbf{r}_2)]$ &
$\Phi^{(c.m.)}_{0,1}(\mathbf{R})
\Phi^\mathrm{(rel)}_{0,2}(\mathbf{r}_{12})$ & 5/16 & 0 & 0 & 1/2 &
11/16
\\[-1.5ex]
3 & & & &
\\[-1.5ex]
& $\psi_{>}^{(3,-)}$ &
$\frac{1}{2}[\sqrt{3}\,u_{1}^{(3,-)}(\mathbf{r}_1,\mathbf{r}_2) +
u_{2}^{(3,-)}(\mathbf{r}_1,\mathbf{r}_2)]$ &
$\Phi^{(c.m.)}_{0,2}(\mathbf{R})
\Phi^\mathrm{(rel)}_{0,1}(\mathbf{r}_{12})$ & 5/16 & 1 & $0,\pm1$
& $1/2,1$ & $11/16,\,3/8$
\\[.5ex]
& $\psi_{>}^{(3,+)}$ &
$\frac{1}{2}[u_{1}^{(3,+)}(\mathbf{r}_1,\mathbf{r}_2) +
\sqrt{3}\,u_{2}^{(3,+)}(\mathbf{r}_1,\mathbf{r}_2)]$ &
$\Phi^{(c.m.)}_{0,3}(\mathbf{R})
\Phi^\mathrm{(rel)}_{0,0}(\mathbf{r}_{12})$ & 5/16 & 0 & 0 & 1/2 &
11/16
\\[1ex]
\hline\hline
\end{tabular}
\end{center}
\end{table*}

The transition to the IP representation can be
done by means of the expansion that holds for $m_{c.m.}, m \ge 0$
(see Appendix \ref{sec:basis-trans})
\begin{eqnarray}
&&\Phi^{(c.m.)}_{0,m_{c.m.}}(\mathbf{R})
\Phi^\mathrm{(rel)}_{0,m}(\mathbf{r}_{12}) = \label{basis-trans}
\\
&&\qquad\qquad \sum_{j=0}^{m_{c.m.}} \sum_{k=0}^m
A_{j,k}^{m_{c.m.},m} \Phi_{0,M-j-k}(\mathbf{r}_1)
\Phi_{0,j+k}(\mathbf{r}_2), \nonumber
\end{eqnarray}
where $M = m_{c.m.} + m$, and
\begin{equation}
A_{j,k}^{m_{c.m.},m} = (-1)^k \bigg(
\begin{matrix}
m_{c.m.} \\ j
\end{matrix}
\bigg) \bigg(
\begin{matrix}
m \\ k
\end{matrix}
\bigg) \sqrt{\frac{(M\!-\!j-\!k)!\,(j+k)!}{2^M\,
m_{c.m.}!\,m!}}, \label{A-coeff}
\end{equation}
combined with the identity $\phi^{(c.m.)}_0(Z)\,
\phi^\mathrm{(rel)}_0(z_{12}) = \phi_0(z_1)\, \phi_0(z_2)$ (in the
3D case). Evidently, at $n_z^{c.m.} = n_z = 0$ the
$z$-component of the orbital function does not contribute to the
entanglement of the full state. Therefore,  without loss of
generality, we can use the 2D model.

If we set $m_{c.m.} = 0$
in Eqs.~(\ref{basis-trans}), (\ref{A-coeff})
( $\Rightarrow$ $j = 0$ and $m = M$),
the following result is obtained for  the states (\ref{low_st0})
(the 2D model) in the IP representation

\begin{eqnarray}
\psi(\mathbf{r}_1,\mathbf{r}_2) &=&
\Phi^{(c.m.)}_{0,0}(\mathbf{R})
\Phi^\mathrm{(rel)}_{0,m}(\mathbf{r}_{12}) \nonumber
\\
&=& \sum_{k=0}^m A_{0,k}^{0,m}\, \Phi_{0,m-k}(\mathbf{r}_1)
\Phi_{0,k}(\mathbf{r}_2). \label{cm0expan}
\end{eqnarray}

At $m = 0$, one has $\psi(\mathbf{r}_1,\mathbf{r}_2) =
\Phi_{0,0}(\mathbf{r}_1) \Phi_{0,0}(\mathbf{r}_2)$, i.e., the
orbital wave function can be written in the form of the product of
wave functions for individual electrons. Since the corresponding
spin state ($S = M_S = 0$) is antisymmetric, the total wave
function has the form of the Slater determinant. Thus, the
entanglement of the lowest state with $m = 0$ must be zero.

At $m=1$, by means of Eq.~(\ref{cm0expan}) we obtain
the antisymmetric function
\beq
\psi(\mathbf{r}_1,\mathbf{r}_2) =
\frac{1}{\sqrt{2}} [\Phi_{0,1}(\mathbf{r}_1)
\Phi_{0,0}(\mathbf{r}_2) - \Phi_{0,0}(\mathbf{r}_1)
\Phi_{0,1}(\mathbf{r}_2)].
\eeq

The corresponding spin state ($S = M_S = 1$) is the
product of individual electron spin states with $m_s = +1/2$.
Evidently, the total wave function is the Slater
determinant, and its entanglement is zero.

At $m \ge 2$ the
expansion (\ref{cm0expan}) contains more than two terms.
The two-electron orbital states are nontrivially correlated, i.e.,
cannot be reduced to the Slater determinant.
The increase of the magnetic quantum number $m$ increases
the number of states in the decomposition (\ref{cm0expan}).
Evidently, it leads to more entangled states.
Finally, the entanglement becomes
maximal (${\cal E} \to 1$) in the limit $m \to \infty$.

The transformation formula (\ref{basis-trans}) demonstrates that
all eigenstates of the Hamiltonian $H$ (including those with
$m_{c.m.} > 0$), that in the limit $V_C \to 0$ converge to
the eigenstates of the noninteracting Hamiltonian $H_0$ with $n_1
= n_2 = 0$, can be written in the form of the product
$\Phi^{(c.m.)}_{0,m_{c.m.}}(\mathbf{R})
\Phi^\mathrm{(rel)}_{0,m}(\mathbf{r}_{12})$  in the same limit
(see Table~\ref{table3}). Since $j+k = m_2$ (and $M-m_2 = m_1$),
we have
\begin{eqnarray}
&&\Phi^{(c.m.)}_{0,m_{c.m.}}(\mathbf{R})
\Phi^\mathrm{(rel)}_{0,m}(\mathbf{r}_{12}) =
\\
&&\qquad\qquad \sum_{j=0}^{m_{c.m.}} \sum_{m_2=j}^{j+m}
A_{j,m_2-j}^{m_{c.m.},m} \Phi_{0,M-m_2}(\mathbf{r}_1)
\Phi_{0,m_2}(\mathbf{r}_2). \nonumber
\end{eqnarray}
By changing the summation order, we obtain
\begin{eqnarray}
&&\Phi^{(c.m.)}_{0,m_{c.m.}}(\mathbf{R})
\Phi^\mathrm{(rel)}_{0,m}(\mathbf{r}_{12}) =
\\
&&\quad\sum_{m_2=0}^M
\sum_{j=\max(0,m_2-m)}^{\min(m_2,m_{c.m.})}
\!\!A_{j,m_2-j}^{m_{c.m.},m} \Phi_{0,M-m_2}(\mathbf{r}_1)
\Phi_{0,m_2}(\mathbf{r}_2). \nonumber
\end{eqnarray}
Comparing this latest relation with Eq.~(\ref{exp-ipb2d0}), one
obtains
\begin{equation}
a_{m_2}^{(-1)^m} =
\sum_{j=\max(0,m_2-m)}^{\min(m_2,m_{c.m.})}
\!\!A_{j,m_2-j}^{m_{c.m.},m}.
\end{equation}
With the aid of this relation and Eqs.(\ref{A-coeff}),
(\ref{trace_orb0}), we determine the trace of the square of the
reduced density matrix (orbital part) and the measure $\cal E$ for
any $m_{c.m.},m \ge 0$. Results for $m_{c.m.},m =
0,1,2,3$ are shown in Table~\ref{table3}. In particular, for
$m_{c.m.} = 0$ (then $j = 0$, $k = m_2$ and $M = m$) one has
$a_{m_2}^{(-1)^m} = A_{0,m_2}^{0,m}$ and
\begin{eqnarray}
\mathrm{Tr}[{\rho_r^\mathrm{(orb)}}^2] = \sum_{k=0}^m
\big|A_{0,k}^{0,m}\big|^4 &\!\!=\!\!& \sum_{k=0}^m \bigg(
\begin{matrix}
 m \\ k
\end{matrix}
\bigg)^{\!\!4} \left[\frac{(m\!-\!k)!\,k!}{2^m\, m!}\right]^2
\nonumber
\\
&\!\!=\!\!& \frac{(2m)!}{(2^m m!)^2}. \label{trace_orb0b}
\end{eqnarray}
This expression and Eq.~(\ref{trspin_lowst}) lead
to the formula (\ref{entm_nint}) for the lowest states with
$m_{c.m.} = 0$ and a given $m$.

The transition from the
IP to the CM representation can be done by the
inverse transformation of Eq.~(\ref{basis-trans}). For $n_1 = n_2
= 0$ and $m_1, m_2 \ge 0$ one has (see Appendix
\ref{sec:basis-trans})
\begin{eqnarray}
&&\Phi_{0,m_1}(\mathbf{r}_1) \Phi_{0,m_2}(\mathbf{r}_2) =
\label{inv-basis-trans}
\\
&&\qquad\qquad \sum_{j=0}^{m_1} \sum_{k=0}^{m_2} A_{j,k}^{m_1,m_2}
\Phi^{(c.m.)}_{0,M-j-k}(\mathbf{R})
\Phi^\mathrm{(rel)}_{0,j+k}(\mathbf{r}_{12}), \nonumber
\end{eqnarray}
where the coefficients $A_{j,k}^{m_1,m_2}$ are given by
Eq.~(\ref{A-coeff}), after replacing $m_{c.m.} \to m_1$, $m
\to m_2$. For example, at $m_1 = m_2 = 1$, one obtains
\bea
&u_2^{(2,+)}(\mathbf{r}_1,\mathbf{r}_2) \equiv
\Phi_{0,1}(\mathbf{r}_1) \Phi_{0,1}(\mathbf{r}_2) =\\
&\frac{1}{\sqrt{2}}[\Phi^{(c.m.)}_{0,2}(\mathbf{R})
\Phi^\mathrm{(rel)}_{0,0}(\mathbf{r}_{12}) -
\Phi^{(c.m.)}_{0,0}(\mathbf{R})
\Phi^\mathrm{(rel)}_{0,2}(\mathbf{r}_{12})]\nonumber.
\eea
This expansion
clearly demonstrates that, although the CM basis functions
formally include correlations between particles (in contrast to
the IP basis functions), non-entangled states (as well as entangled)  
states (such as $u_2^{(2,+)}$)
can be regularly represented in this basis.

Regarding the last remarks, we point out that an analysis, if
performed strictly in the CM representation, may lead in some
cases to incorrect conclusions. An example is the analysis of the
role of the inter-particle interaction in preparing entangled states.
Namely, the form of the lowest states in the IP representation is
nontrivially determined by the electron-electron interaction $V_C
= \lambda/r_{12}$, even in the limit $\lambda \to 0$. In this
limit the values of coefficients in the superposition
(\ref{superposition}) are specified (see Sec.~\ref{sec:indiv_pd}).
Since the coefficients are not arbitrary (in contrast to the case
of the exact $\lambda = 0$), the states are entangled (see
Table~\ref{table3}). In the CM representation, however, the same
states are simply the eigenstates of $H_0$ with specified (c.m.)
symmetries. These symmetries exist regardless of whether the
interaction $V_C$ is present or not (due to decoupling of the c.m.
and relative motions). Thus, in this representation the
eigenstates of $H$ (in the limit $V_C \to 0$) and $H_0$ are the
same. One might conclude that the interaction $V_C$ is not
necessary to prepare an entangled state. Thus, we have obtained
two opposite conclusions by analyzing the same effect in different
representations.

The solution of this paradox lies in the fact that in the limit
$V_C \to 0$ the form of interaction $V_C$ is not crucial -- the
only requirement is that it must be a function of
the relative distance $\mathbf{r}_{12}$ only (i.e., it must be independent of
$\mathbf{R}$) in order to keep the c.m. and relative motions
decoupled. Then the interaction breaks only the symmetries of the
noninteracting system, that are related to the integrals of motion
of individual particles. This symmetry breaking removes the
degeneracy (see the last paragraph in
Sec.~\ref{sec:int-of-motion}). It reduces the set of all
eigenstates of $H_0$ to the subset of those that have 
symmetries related to the c.m. integrals of motion. Consequently,
just this subset of eigenstates of $H_0$ is the complete set of
eigenstates of $H$ in the limit $V_C \to 0$. These states,
although being the CM basis states,
have the form of certain linear combinations of the
IP basis states, i.e., they are entangled.
The inter-particle interaction selects the eigenstates of a proper
symmetry at the limit $\lambda \to 0$.

\section{The entanglement of the ground state in the magnetic field }
\label{sec:ground_st}
\begin{figure}
\vspace{-.1in} \epsfxsize 2.6in \epsffile{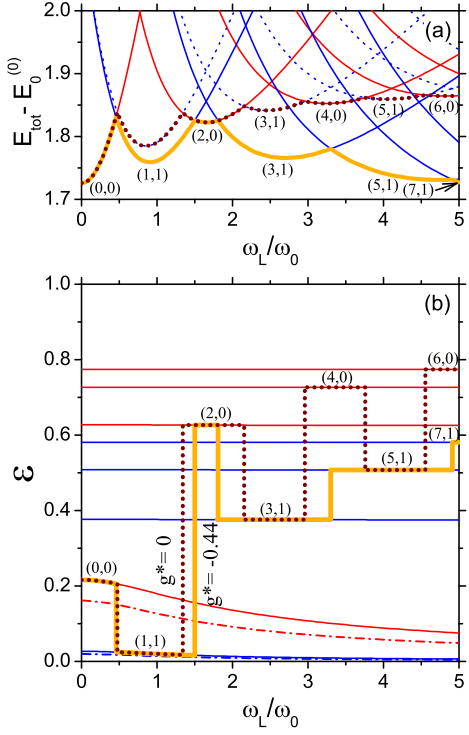}
\caption{(Color online) (a) Lowest energy levels (thin red and
blue lines) and the ground state energy (thick lines) of axially
symmetric two-electron QDs with $\lambda = 2$ (in
$\hbar\omega_0\ell_0$ units) for $\omega_z/\omega_0 \gg 1$. In
order to get a better resolution, the energies are defined with
respect to the $E_0^{(0)}$ level (all energies are defined
in $\hbar\omega_0$ units). The results for $g^* = 0$ and
$-0.44$ are represented by dotted and solid lines, respectively.
The numbers in parentheses are the values of quantum numbers $(m,S)$
that characterize the corresponding states. b) Entanglement of
the states corresponding to the levels shown in (a). Thick
dotted and solid (orange) lines represent the entanglement of the
ground state at $g^* = 0$ and $-0.44$, respectively. Dash-dotted
lines show the entanglement measure of the lowest levels with $M = 0$
and $1$ obtained within the first-order approximation (see
Eqs.~(\ref{ent-1st-order}), (\ref{rel-dE-b1})).}
\label{fig:entgr2d}
\end{figure}

At $V_C = 0$ the ground state orbital wave function with $m = 0$
is the product
\beq \psi^{(0)}_\mathrm{gr} =
\Phi^{(c.m.)}_{0,0}\,\phi^{(c.m.)}_0\,
\Phi^\mathrm{(rel)}_{0,0}\,\phi^\mathrm{(rel)}_0 \eeq
(or $\Phi^{(c.m.)}_{0,0} \Phi^\mathrm{(rel)}_{0,0}$ within
the 2D model). Among the states (\ref{low_st0}), it
has the lowest energy at all values of the magnetic field.

The electron-electron interaction does not affect
the c.m. motion.  Therefore, even at $V_C \neq 0$, the ground state is
characterized by $m_{c.m.} = 0$ for all values of the
magnetic field $B$ and the electron-electron interaction strength
$\lambda$. The interaction leads, however, to the coupling
(\ref{psi_rm}) in $\psi_\mathrm{rel}$.  This results in
the crossings of the energy levels (as functions of $B$)
with different values of the quantum number $m$ [see
Fig.~\ref{fig:entgr2d}(a)].
 In addition, if we take into account the Zeeman splitting,
for a negative Land\'e factor $g^* < 0$ the ground state spin
quantum number of the $m$-th segment is determined by
Eq.~(\ref{M_S}). According to this formula, the total spin $S
=\frac{1}{2}[1 - (-1)^m]$ alternates between $0$ and $1$. This
effect leads to the well known spin oscillations or
singlet-triplet (ST) transitions in the ground state (see
Refs.~\onlinecite{rev1,rev2} for a review).

As an example, the evolution of the ground state energy
for two values of the Land\'e factor is shown on
Fig.~\ref{fig:entgr2d}(a).  A typical effect due to the Zeeman
splitting is the dilation of the triplet state segments ($S = M_S
= 1$) at the cost of the singlet ones ($S = 0$) with an increase
of the magnetic field. The levels, that correspond to the singlet
states, do not depend on $g^*$. The triplet states are lower for
$g^* = -0.44$ than for $g^* = 0$. As a consequence, in the case
$g^* = -0.44$ the segments $(m,S) = (0,0)$ and especially $(2,0)$
are reduced, and for $m > 2$ the singlet ground states are fully
suppressed.

The measure $\cal E$,  calculated for each
ground state segment [Fig.~\ref{fig:entgr2d}(a)]
separately, yields
a  discontinues function ${\cal E}_\mathrm{gr}(B)$
 of the magnetic field [see Fig.~\ref{fig:entgr2d}(b)].
 As we have seen in Fig.~\ref{fig:entm_nint},
 the entanglement of the
lowest state with $m=0$ decreases by increasing the magnetic
field. For the magnetic number $m \ge 2$ this dependence is, however,  very weak
[see Fig.~\ref{fig:entgr2d}(b)].  In fact, for these states the measure
$\cal E$ can be approximated by constant values that are close to the
values (\ref{entm_nint}) for the noninteracting case.

The Zeeman splitting does not affect the orbital wave function, and,
therefore, the measure. However, different values of
the Land\'e factor determine the position and length
of $(m,S)$-segments [see Fig.~\ref{fig:entgr2d}(a)].
As a result, the variation of the measure ${\cal E}_\mathrm{gr}(B)$
 will be different for different values
of $g^*$ [see Fig.~\ref{fig:entgr2d}(b)].

By increasing the magnetic
field, the ST transitions occur periodically
at $g^* = 0$.
Then, ${\cal E}_\mathrm{gr}(B)$ is the combination [see
Eq.~(\ref{ent_measure})] of a "square-wave" function
$\mathrm{Tr}[{\rho_r^\mathrm{(spin)}}^2]$ and a step function
$\mathrm{Tr}[{\rho_r^\mathrm{(orb)}}^2]$. On the other hand, for
$g^* = -0.44$ the ST transitions appear only a few times at lower
values of the field. Finally, the system settles down in the triplet ground state
with the quantum numbers $S = M_S = 1$. Consequently, the function ${\cal
E}_\mathrm{gr}(B)$ performs only a few oscillations in this case
and, after that, it has the step function form.

\begin{figure}
\vspace{-.1in} \epsfxsize 2.6in \epsffile{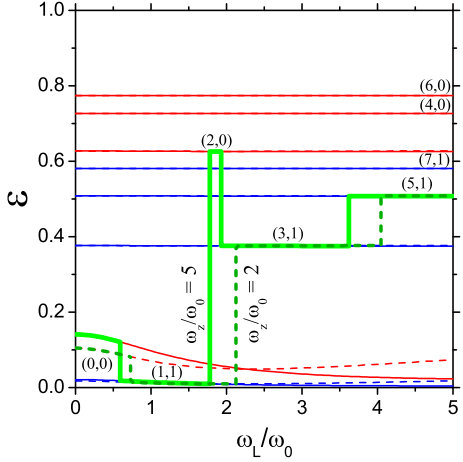}
\caption{(Color online) Entanglement of the lowest states
with different $m$ (thin red and blue lines) of axially symmetric
two-electron QDs with $\lambda = 2$ (in $\hbar\omega_0\ell_0$
units), $g^* = -0.44$ and two values of the ratio
$\omega_z/\omega_0$ equal $2$ (dashed lines) and $5$ (solid lines).
The numbers in parentheses are the values of quantum numbers $(m,S)$
which characterize the corresponding states. Entanglement of the
ground state for these two cases is represented by the thick green
line (dashed and solid, respectively).} \label{fig:entgr3d}
\end{figure}

Figure~\ref{fig:entgr3d} shows the measure ${\cal E}(B)$ of the
lowest states that contribute to the ground state, as well as
${\cal E}_\mathrm{gr}(B)$, for two different values of the ratio
$\omega_z/\omega_0$ and for $g^* = -0.44$. For larger values of
the ratio $\omega_z/\omega_0$ the function ${\cal
E}_\mathrm{gr}(B)$ is similar to that obtained in the limit
$\omega_z/\omega_0 \to \infty$ [shown in
Fig.~\ref{fig:entgr2d}(b)]. By decreasing
$\omega_z/\omega_0$, the width of the peak related to the state
$(2,0)$ becomes smaller, and the positions of discontinuities are
shifted to higher values of the magnetic field (see the results
for $\omega_z/\omega_0 = 5$ in Fig.~\ref{fig:entgr3d}).
At a sufficiently small value of the ratio
$\omega_z/\omega_0$ the peak $(2,0)$ disappears (see the case
$\omega_z/\omega_0 = 2$ in Fig.~\ref{fig:entgr3d}). The function
${\cal E}_\mathrm{gr}(B)$ has the step function form, except at
the beginning [the segment $(0,0)$]. Evidently, the
variation of this form with the change of the ratio
$\omega_z/\omega_0$ is due to the shift of the ST-transition
points \cite{sc}, since  for a given $m$ the function ${\cal
E}(B)$ does not change significantly.

\section{Estimation of the entanglement from
measurable quantities}
\label{exme}

From our analysis (Sec.~\ref{sec:CM-rep}) it follows that the
entanglement of the lowest states with $m \ge 2$ depends only
weakly on the strength of the Coulomb interaction $\lambda$ (see
Fig.~\ref{fig:ent-rw}). It is also found that the entanglement of
these states is weakly dependent on the ratio $\omega_z/\omega_0$,
and on the magnetic field strength (see Fig.~\ref{fig:ent-oml}).
These results suggest that Eq.~(\ref{entm_nint}), that determines
(exactly) the entanglement measure $\cal E$ of the lowest states
in the limit $\lambda \to 0$, can be used as the zeroth-order
approximation in the general case. In other words, formula
(\ref{entm_nint}) enables us to estimate the entanglement of the
ground state of a QD with  interacting electrons at the
values of the magnetic field when $m \ge 2$. The only information
required is just the value of the quantum number $m$ (or $M$,
because $m_{c.m.} = 0$). This feature is a consequence of the
chosen (parabolic and axially symmetric) form of the confining
potential. The magnetic quantum number $M$ can be determined
easily from the magnetic field dependence of the ground state
energy and positions of the singlet-triplet transitions.

The zeroth-order approximation fails, however, if $M = 0$ or 1,
since for these two values Eq.~(\ref{entm_nint}) gives
${\cal E} = 0$.  The valuable results can be obtained within the first-order
approximation, if we set $n_\mathrm{max} = 1$ (and $n_z^\mathrm{max} =
0$) in the expansion (\ref{trace3d}). As a result, we have
\begin{equation}
{\cal E} = 1 - \frac{3-(-1)^m}{2} \Big[ I_0\, b_0^4 + I_1 \Big(4\,
b_0^2 \!+ b_1^2\Big) b_1^2 \Big]\,. \label{ent-1st-order}
\end{equation}
Here, for the sake of convenience, we introduce the notations:
$I_0 = I(0,0,0,0;m)$, $I_1 = I(0,0,1,1;m) \equiv I(1,1,1,1;m)$,
and $b_n = b_{n,0}^{(m)}$. The values of $I_0$ are defined by
Eq.~(\ref{I0}), whereas $I_1 = 1/4$, $3/16$, $5/32$, $35/256$ for
$m = 0$, $1$, $2$, $3$, respectively. To apply formula
(\ref{ent-1st-order}), in addition to the quantum number $m$, one
needs to know also the coefficients $b_0$ and $b_1$. These
coefficients depend on the QD parameters and on the strength of
the magnetic field ($\omega_0$, $\lambda$ and $\omega_L$). Below we
suggest how to estimate the coefficients $b_0$ and
$b_1$ from the magnetic field dependence of the QD ground state
energy.

Starting from the Hamiltonian for the relative motion (\ref{relham})
and applying the Hellmann-Feynman theorem, one obtains
\begin{equation}
\frac{\partial E_\mathrm{rel}}{\partial\omega_L} =
\mu\;\!\omega_L\langle\rho_{12}^2\rangle - \langle
l_z^\mathrm{rel}\rangle.
\label{der}
\end{equation}
Since we consider the QD in the ground state (with a given $M =
m$), one has
\begin{eqnarray}
&&\langle l_z^\mathrm{rel}\rangle = \hbar m\,,\\
&& \langle\rho_{12}^2\rangle \approx \frac{\hbar}{\mu\Omega} (m +
1 - 2 b_0 b_1 \sqrt{m\!+\!1} + 2 b_1^2),
\end{eqnarray}
where
\begin{equation}
b_0 \approx (1 - b_1^2)^{1/2}\, \label{b0}
\end{equation}
(see Appendix \ref{sec:meanv-rho2}). As a result,
Eq.~(\ref{der}) transforms to the form
\begin{equation}
\frac{\partial E_\mathrm{rel}}{\partial\omega_L} \approx
\hbar\frac{\omega_L}{\Omega} \Big(m + 1 - 2\sqrt{m+1}\sqrt{b_1^2(1
- b_1^2)} + 2 b_1^2\Big) - \hbar m\,.
\label{rel-dE-b1}
\end{equation}
By means of elementary algebraic transformations, one
obtains
\begin{equation}
(m+2)\,b_1^4 + (F-m-1)\,b_1^2 + F^2/4 = 0, \label{biquad-eq}
\end{equation}
where
\begin{equation}
F(\omega_L) = 1 + \bigg(1-\frac{\Omega}{\omega_L} \bigg)\,m -
\frac{\Omega}{\omega_L} \frac{1}{\hbar}\frac{\partial
E_\mathrm{rel}}{\partial\omega_L}.
\end{equation}
If at a given magnetic field strength B ($\sim\omega_L$) the values of
the magnetic quantum number $m$ and $\partial
E_\mathrm{rel}/\partial\omega_L$ are known, one can evaluate the
expansion coefficients $b_1$ and $b_0$ with the aid of
Eqs.~(\ref{biquad-eq}), (\ref{b0}). By virtue of
Eq.~(\ref{ent-1st-order}) it is straightforward to determine
approximately the entanglement of the ground state of the QD at a
given value of the magnetic field.

To illustrate the proposal we calculate the measure $\cal E$ for
the ground state energy as a function of the magnetic field, shown
in Fig.~\ref{fig:entgr2d}(a). The obtained values of the measure
$\cal E$ for $M = 0$ and $1$ reproduce qualitatively the exact
dependence of the measure ${\cal E}(\omega_L)$, shown in
Fig.~\ref{fig:entgr2d}(b) (the deviation is less than $30\%$). The
results for the states $M \ge 2$ are in a remarkable agreement
with exact values (in Fig.~\ref{fig:entgr2d}(b) they
practically coincide).
For example, at the value $\omega_L/\omega_0 = 1.65$ the ground
state is characterized by $M = 2$ and $S = M_S = 0$. This value
belongs to the narrow segment between the second and  third ST
transitions shown in Fig.~\ref{fig:entgr2d}. [This state is also
marked in Figs.~\ref{fig:low_levels}(b) and \ref{fig:c1c2}(b,c) by
a red closed circle -- the lower one.] Its entanglement measure is ${\cal
E} = 0.6265$ (the exact value), whereas the values within the zeroth-
and first-order approximation are ${\cal E}^{(0)} = 0.625$ and
${\cal E}^{(1)} = 0.6261$, respectively.

We point out that the relation (\ref{rel-dE-b1})  is crucial
[together with Eq.~(\ref{ent-1st-order})] for estimation of the
entanglement of the QD ground state from experimental data. In
fact, one has to know only the  addition energy $E_a$ for two-electron QDs
\begin{equation}
E_a = \mu(2) - \mu(1) = E_\mathrm{tot}(2) - 2E_\mathrm{tot}(1)
\end{equation}
as a function of the magnetic field.
Here $\mu(N) = E_\mathrm{tot}(N) - E_\mathrm{tot}(N-1)$ and
$E_\mathrm{tot}(N)$ are the chemical potential and the total
ground state energy of the QD with $N$ electrons, respectively.
Thus, one has for one and two electrons
\begin{eqnarray}
&&E_\mathrm{tot}(1) = \hbar\Omega + \frac{\hbar\omega_z}{2} + g^* \mu_B
B m_s,
\\
&&E_\mathrm{tot}(2) = E_\mathrm{rel} + E_\textsc{cm} + g^* \mu_B B
M_s,
\end{eqnarray}
where $E_{c.m.} = \hbar\Omega + \hbar\omega_z/2$. Assuming that in
the one-electron ground state the spin projection $m_s = -1/2$, and applying the
relation $\mu_B B = (m^*/m_e)\hbar\omega_L$, we obtain
\begin{equation}
E_a = E_\mathrm{rel} - \bigg(\hbar\Omega +
\frac{\hbar\omega_z}{2}\bigg) + g^* \frac{m^*}{m_e}\,
\hbar\omega_L (M_s+1).
\end{equation}
As a result, the function $F(\omega_L)$ is expressed in terms of
the derivative $\partial E_a/\partial\omega_L$
\begin{equation}
F = \bigg(1-\frac{\Omega}{\omega_L} \bigg)\,m +
\frac{\Omega}{\omega_L} \bigg[ g^*\frac{m^*}{m_e}\,(M_S+1) -
\frac{1}{\hbar}\frac{\partial E_a}{\partial\omega_L} \bigg].
\label{F_Ea}
\end{equation}
If one knows the evolution of the  addition energy  $E_a$ in the
magnetic field, it becomes possible to evaluate $\partial
E_a/\partial\omega_L$ at different values of $\omega_L$ and
determine the function $F(\omega_L)$. Evidently, it will be a
discontinuous function at those values of $\omega_L$,
where the ground state changes the ($m,M_S$)-values
(singlet-triplet transitions). However, since $F(\omega_L)$
usually changes slowly in intervals between two transition points,
it is sufficient to determine the value of $F$ at an arbitrary
point in each interval. It is particularly convenient if $E_a$ has
a local minimum at a $\omega_L$-value inside the interval. At this
point $\partial E_a/\partial \omega_L = 0$, and one needs to know
only the corresponding value of $\omega_L$ (or $B$) and the
magnetic quantum number $m$ to evaluate Eq.~(\ref{F_Ea}).

The proposed method is founded on the assumption that
axially symmetric two-electron QDs
can be approximated by the parabolic model.
The ground state energy of the dot is calculated assuming that the dot is isolated.
This approximation is well justified, when the tunneling between the QD
and an external source and drain is relatively weak.
We do not take into account the effect of finite temperature; this is appropriate
for experiments which are performed at temperatures $k_BT\ll \hbar\omega_0$,
with $\hbar\omega_0\sim 2 - 5$ meV being the mean level spacing.
Note that for these experiments a typical temperature is estimated
to be below $100$ mK $(0.008$ meV) \cite{kov,ni2}.

The analysis of numerous experimental data confirms that
an effective trapping potential in small QDs with a few electrons
is quite well approximated by a parabolic confinement \cite{man,sar,rev2}.
Indeed, the validity of this approximation has been proven by the
observation of the {\it shell structure} of a parabolic potential in small
vertical quantum dots \cite{kov,ni2},
that was predicted theoretically
in a number of publications \cite{sh1,sh2,sh3}.
Furthermore, a good agreement between experimental data
and theoretical calculations
of the addition energy has been demonstrated
within the parabolic model as well (for a review see \cite{rev2}).
Taking into account the rapid development of the
nano-sized technology and measurement techniques in the last
decade we are very optimistic that our method could be useful
to trace the entanglement properties in two-electron quantum dots
in the magnetic field.

\section{Summary}
\label{sec:conclusions}

We have considered a QD model that consists of two
interacting electrons, confined in the axially-symmetric parabolic
potential, in the external magnetic field. Within
this model we have investigated in detail the connection between
orbital correlations and the entanglement of the lowest
two-electron states. These states compose a set of ground
states at various values of the magnetic field.

We have analyzed
integrals of motion and symmetries at zero and nonzero
electron-electron interaction $V_C$. This analysis enabled us
to introduce two appropriate basis sets: (i) the IP basis the
elements of which are common eigenstates of the orbital part of the
Hamiltonian with noninteracting electrons ($H_0$) and of the
integrals of motion of individual electrons; (ii) the CM basis
the elements of which are common eigenstates of the Hamiltonian $H_0$ and
of the integrals of the c.m. and relative motions. To quantify the
entanglement of two-electron states we have used the measure based
on the linear entropy.
With the aid of these representations we
investigated how the orbital entanglement evolves at tuning of the
interaction and the magnetic field strengths. For this purpose
we have developed an analytical approach that enables to us
to calculate the entanglement
measure for interacting and noninteracting electrons.

We have established the connection between the IP and CM
representations of two-electron states, that allows us to readily 
illuminate quantum correlations. For example, in the limit
of noninteracting electrons ($V_C \to 0$) in the CM representation
the correlations between individual particles are hidden.
Therefore, it is not quite evident whether a given state is
entangled or not. To determine the entanglement one has either to
calculate the entanglement measure, or to make a transition to the
IP representation.
We have shown that the elements of the CM basis are formally
entangled (i.e., include inter-particle correlations), even at zero
interaction. This fact,
however, does not imply that the stationary states of two
non-interacting electrons are entangled. Due to the degeneracy of the
energy levels at $V_C = 0$, one can always choose a set of
eigenstates that exhibit a zero entanglement. Such a set is here the
IP basis.

By means of our findings, we have studied the entanglement of the ground states of
two-electron QDs in the magnetic field. It was demonstrated that at the
magnetic quantum number $M\geq2$ of the ground state the
entanglement measure is remarkably well reproduced by means of
analytical expressions obtained in the limit of noninteracting
electrons $V_C \to 0$. Our analysis predicts a nonhomogeneous behavior
of the entanglement as a function of the magnetic field. This
feature arises due to singlet-triplet transitions. As soon as the
singlet states are suppressed by the magnetic field, the entanglement
is growing as a step function with an increase of the magnetic
quantum number $M$.

By virtue of our analysis, we have proposed a practical approach
to  trace the evolution  of the entanglement with the aid of the
addition energy of two-electron QDs in the magnetic field.
We hope that this approach would provide a practical method
to measure  the entangled states of QDs in the magnetic field.

\section*{Acknowledgments}
This work was supported in part by RFBR Grant 14-02-00723 and by
the Project 171020 of Ministry of Education and Science of Serbia.
N.S.S. is grateful for the warm hospitality at UIB.

\appendix
\section{The individual particle basis}
\label{sec:IP-basis}

The IP basis elements are products of eigenstates of the Hamiltonians
$H_1$ and $H_2$. A favourable set of eigenstates of the
single-electron Hamiltonians $H_i$ ($i = 1,2$) within the 2D model
are the Fock-Darwin states $\Phi_{n_i,m_i}(\mathbf{r}_i)$, where
$n_i = 0, 1, 2, \ldots$ and $m_i = 0, \pm1, \pm2, \ldots$ are the
radial and magnetic quantum number of the electrons. The IP basis
elements in the 2D model are, therefore, the products
$\Phi_{n_1,m_1}(\mathbf{r}_1)\, \Phi_{n_2,m_2}(\mathbf{r}_2)$.

The Fock-Darwin states are stationary states of a charged
particle, confined in an axially symmetric 2D parabolic potential,
in a perpendicular magnetic field. Their explicit form is
\cite{FD}
\begin{eqnarray}
\Phi_{n_i,m_i}(\rho_i,\varphi_i) \!\!&=&\!\!
\sqrt{\frac{\bar{\Omega}}{\pi}} \sqrt{\frac{n_i!}{(n_i+|m_i|)!}}\,
\big(\sqrt{\bar{\Omega}}\, \rho_i\big)^{|m_i|} \nonumber
\\
&& e^{-\frac{1}{2}\bar{\Omega}\, \rho_i^2}\,
\mathrm{L}_{n_i}^{|m_i|}(\bar{\Omega}\, \rho_i^2)\,
e^{im_i\varphi_i}, \label{fd-states}
\end{eqnarray}
where $\rho_i = (x_i^2 + y_i^2)^{1/2}$, $\varphi_i =
\arctan(y_i/x_i)$, $\bar{\Omega} = m^* \Omega/\hbar =
(\Omega/\omega_0)/\ell_0^2$ and $\mathrm{L}_{n_i}^{|m_i|}$ are the
Laguerre polynomials. These states are simultaneously the
eigenstates of $l_z^{(i)}$. The parity of the states is
$(-1)^{m_i}$ and the corresponding (Fock-Darwin) energies are
\begin{equation}
E_{n_i,m_i} = \hbar\Omega(2n_i + |m_i| + 1) - \hbar\omega_L
m_i.\label{fd_levels}
\end{equation}

A set of eigenstates of the single-electron Hamiltonians $H_i$ ($i
= 1,2$) in the 3D model are the products
$\Phi_{n_i,m_i}(\rho_i,\varphi_i)\, \phi_{n_{zi}}(z_i)$, where
\begin{equation}
\phi_{n_{zi}}(z_i) =
\frac{(\bar{\omega}_z/\pi)^{1/4}}{\sqrt{2^{n_{zi}}n_{zi}!}}\,
e^{-\frac{1}{2}\bar{\omega}_z z_i^2}\,
\mathrm{H}_{n_{zi}}(\sqrt{\bar{\omega}_z}\,z_i), \label{z_states}
\end{equation}
are the eigenfunctions of the harmonic oscillator in the
$z$-direction. Here $\bar{\omega}_z = m^* \omega_z/\hbar =
1/\ell_z^2$\,, and $\mathrm{H}_{n_{zi}}$ are the Hermite
polynomials ($n_{zi} = 0, 1, 2, \ldots$). The parity of the states
$\Phi_{n_i,m_i}(\rho_i,\varphi_i)\, \phi_{n_{zi}}(z_i)$ is
$(-1)^{m_i+n_{zi}}$. The related energy levels are
\begin{equation}
E_{n_i,m_i,n_{zi}} = E_{n_i,m_i} + \hbar\omega_z(n_{zi} +
\hbox{$\frac{1}{2}$}). \label{H0lev3D}
\end{equation}
Therefore, the IP basis elements in the 3D model are the products
$\Phi_{n_1,m_1}(\mathbf{r}_1)\, \phi_{n_{z1}}(z_1)\,
\Phi_{n_2,m_2}(\mathbf{r}_2)\, \phi_{n_{z2}}(z_2)$.

The (anti)symmetrized products of the Fock-Darwin states are
defined as
\begin{eqnarray}
\{\Phi_{n_1,m_1}(\mathbf{r}_1), \Phi_{n_2,m_2}(\mathbf{r}_2)\}_\pm
\equiv \qquad\qquad && \nonumber
\\
A [\Phi_{n_1,m_1}(\mathbf{r}_1) \Phi_{n_2,m_2}(\mathbf{r}_2) \pm
\Phi_{n_2,m_2}(\mathbf{r}_1) \Phi_{n_1,m_1}(\mathbf{r}_2)],
&&\label{ipb2d-sym}
\end{eqnarray}
were $A = 1/2$ if $n_1 = n_2$ and $m_1 = m_2$, otherwise $A =
1/\sqrt{2}$. Note that $\{\Phi_{n_1,m_1}(\mathbf{r}_1),
\Phi_{n_2,m_2}(\mathbf{r}_2)\}_{-} = 0$ if $n_1 = n_2$ and $m_1 =
m_2$. The states (\ref{ipb2d-sym}) are the elements of the
(anti)symmetric counterparts of the IP basis in the 2D model. As
the unsymmetrized states (\ref{fd-states}), they are related to
the eigenenergies (\ref{fd_levels}).

\section{The center-of-mass basis}
\label{sec:CM-basis}

The CM basis elements are products of eigenstates of
$H_{c.m.}$ and $H^{(0)}_\mathrm{rel}$. These states have the
same form as the single-electron states, but with the values
${\cal M} = 2m^*$ and $\mu = m^*/2$, respectively, instead of the
effective electron mass $m^*$.

Therefore, the CM basis elements in the 2D model are the products
$\Phi_{n_{c.m.}, m_{c.m.}}^{(c.m.)} (\mathbf{R})\,
\Phi_{n,m}^\mathrm{(rel)} (\mathbf{r}_{12})$, where
\begin{eqnarray}
&&\Phi_{n_{c.m.}, m_{c.m.}}^{(c.m.)}
(\rho_{c.m.},\varphi_{c.m.}) =
\sqrt{\frac{2\bar{\Omega}}{\pi}}
\sqrt{\frac{n_{c.m.}!}{(n_{c.m.}+|m_{c.m.}|)!}}
\nonumber
\\
&&\big(\sqrt{2\bar{\Omega}}\,
\rho_{c.m.}\big)^{|m_{c.m.}|}\, e^{-\bar{\Omega}
\rho_{c.m.}^2}\,
\mathrm{L}_{n_{c.m.}}^{|m_{c.m.}|}(2\bar{\Omega}\,
\rho_{c.m.}^2)\times\nonumber\\
&& \times e^{im_{c.m.}\varphi_{c.m.}}
\label{cmfd-states}
\end{eqnarray}
and
\begin{eqnarray}
\Phi_{n, m}^\mathrm{(rel)} (\rho,\varphi) \!\!&=&\!\!
\sqrt{\frac{\bar{\Omega}}{2\pi}} \sqrt{\frac{n!}{(n+|m|)!}}\,
\bigg(\sqrt{\frac{\bar{\Omega}}{2}}\, \rho\bigg)^{|m|} \nonumber
\\
&& e^{-\frac{1}{4}\bar{\Omega}\, \rho^2}\,
\mathrm{L}_n^{|m|}(\bar{\Omega}\, \rho^2/2)\, e^{im\varphi}.
\label{relfd-states}
\end{eqnarray}
Here $\rho_{c.m.} = (X^2 + Y^2)^{1/2}$, $\varphi_{c.m.}
= \arctan(Y/X)$, $\rho = (x^2 + y^2)^{1/2}$ and $\varphi =
\arctan(y/x)$.
The labels $n_{c.m.},n = 0, 1, 2, \ldots$ and
$m_{c.m.},m = 0, \pm1, \pm2, \ldots$ are the radial and
magnetic quantum number of the c.m. and the relative motion degree
of freedom, respectively.
The states (\ref{cmfd-states}) and (\ref{relfd-states}) are
simultaneously the eigenstates of $l_z^{(c.m.)}$ and
$l_z^{\mathrm{(rel)}}$, respectively.
The parities of these states are $(-1)^{m_{c.m.}}$ and
$(-1)^m$, whereas the corresponding (Fock-Darwin) energies are
\begin{eqnarray}
E_{n_{c.m.},m_{c.m.}} &=& \hbar\Omega(2n_{c.m.} +
|m_{c.m.}| + 1) - \hbar\omega_L m_{c.m.},
\label{cmfd_levels}
\\[.5ex]
E_{n,m} &=& \hbar\Omega(2n + |m| + 1) - \hbar\omega_L m.
\label{relfd_levels}
\end{eqnarray}

The CM basis elements in the 3D model are $\Phi_{n_{c.m.},
m_{c.m.}}^{(c.m.)}
(\rho_{c.m.},\varphi_{c.m.})\,
\phi_{n_z^{c.m.}}^{(c.m.)}(Z)\,
\Phi_{n,m}^\mathrm{(rel)}(\rho_{12},\varphi_{12})\,
\phi_{n_z}^\mathrm{(rel)}(z_{12})$, where
\begin{equation}
\phi_{n_z^{c.m.}}^{(c.m.)}(z_{c.m.}) =
\frac{(2\bar{\omega}_z/\pi)^{1/4}}{\sqrt{2^{n_z^{c.m.}}n_z^{c.m.}!}}\,
e^{-\bar{\omega}_z z_{c.m.}^2}\,
\mathrm{H}_{n_z^{c.m.}}(\sqrt{2\bar{\omega}_z}\,z_{c.m.})
\label{cmz_states}
\end{equation}
and
\begin{equation}
\phi_{n_z}^\mathrm{(rel)}(z) =
\frac{(\bar{\omega}_z/2\pi)^{1/4}}{\sqrt{2^{n_z}n_z!}}\,
e^{-\frac{1}{4}\bar{\omega}_z z^2}\,
\mathrm{H}_{n_z}(\sqrt{\bar{\omega}_z/2}\,z), \label{relz_states}
\end{equation}
($n_z^{c.m.},n_z = 0, 1, 2, \ldots$). The parities of the
states $\Phi_{n_{c.m.}, m_{c.m.}}^{(c.m.)}
(\rho_{c.m.},\varphi_{c.m.})\,
\phi_{n_z^{c.m.}}^{(c.m.)}(Z)$ and
$\Phi_{n,m}^\mathrm{(rel)}(\rho,\varphi)\,
\phi_{n_z}^\mathrm{(rel)}(z)$ are
$(-1)^{m_{c.m.}+n_z^{c.m.}}$ and $(-1)^{m+n_z}$,
respectively. The related eigenenergies are
\begin{eqnarray}
E_{n_{c.m.},m_{c.m.},n_z^{c.m.}} &=&
E_{n_{c.m.},m_{c.m.}} + \hbar\omega_z(n_z^{c.m.} +
\hbox{$\frac{1}{2}$}), \label{Hcmlev3D2}
\\[.5ex]
E_{n,m,n_z} &=& E_{n,m} + \hbar\omega_z(n_z +
\hbox{$\frac{1}{2}$}). \label{H0lev3D2}
\end{eqnarray}

The CM basis functions have a definite exchange symmetry by
construction. Indeed, the exchange of the particles 1 and 2 is
equivalent to the transformation $\mathbf{r}_{12} \to
-\mathbf{r}_{12}$ and the exchange symmetry of $\psi
(\mathbf{r}_1, \mathbf{r}_2)$ is directly determined by the parity
of $\psi_\mathrm{rel} (\mathbf{r}_{12})$ and {\it vice versa}.
Thus, the orbital wave function is symmetric (antisymmetric) if
$\psi_\mathrm{rel} (\mathbf{r}_{12})$ is even (odd). The c.m.
coordinate $\mathbf{R}$ is not affected by this transformation. As
a result, the wave function $\psi_{c.m.}$ does not change the
sign by exchanging the particles. Therefore, if we construct the
orbital wave function as the product (\ref{psiorb}), where the
relative wave function is defined by (\ref{psi_rm}) (with a fixed
parity of the index $n_z$ in the 3D case), the Pauli principle
will be encountered automatically.

\section{The $I$ and $J$ integrals}
\label{sec:IJ-integrals}

The $I$ and $J$ integrals appearing in the expansion
(\ref{trace3d}) are
\begin{eqnarray}
&&I(n_1,n_2,n_3,n_4;m) = \int \cdots \int d\mathbf{r}_1\,
d\mathbf{r}_1^{\,\prime}\, d\mathbf{r}_2\,
d\mathbf{r}_2^{\,\prime} \times \nonumber
\\
&&\qquad\qquad \Phi_{0,0}^{(c.m.)}(\hbox{$\frac{\mathbf{r}_1
+ \mathbf{r}_2}{2}$})\,
{\Phi_{0,0}^{(c.m.)}}^*(\hbox{$\frac{\mathbf{r}_1^{\,\prime}
+ \mathbf{r}_2}{2}$}) \times \nonumber
\\
&&\qquad\qquad
{\Phi_{0,0}^{(c.m.)}}^*(\hbox{$\frac{\mathbf{r}_1 +
\mathbf{r}_2^{\,\prime}}{2}$})\,
\Phi_{0,0}^{(c.m.)}(\hbox{$\frac{\mathbf{r}_1^{\,\prime} +
\mathbf{r}_2^{\,\prime}}{2}$}) \times \nonumber
\\
&&\qquad\qquad \Phi^\mathrm{(rel)}_{n_1,m}(\mathbf{r}_1 \!\!-\!
\mathbf{r}_2)\,
{\Phi^\mathrm{(rel)}_{n_2,m}}^{\!\!*}(\mathbf{r}_1^{\,\prime}
\!\!-\! \mathbf{r}_2) \times \nonumber
\\
&&\qquad\qquad {\Phi^\mathrm{(rel)}_{n_3,m}}^{\!\!*}(\mathbf{r}_1
\!\!-\!\mathbf{r}_2^{\,\prime})\,
\Phi^\mathrm{(rel)}_{n_4,m}(\mathbf{r}_1^{\,\prime} \!\!-\!
\mathbf{r}_2^{\,\prime}) \label{int_I_def}
\end{eqnarray}
(here $\mathbf{r}_i$ are vectors in the $xy$-plane) and
\begin{eqnarray}
&&J(n_{z_1},n_{z_2},n_{z_3},n_{z_4}) = \int \cdots \int dz_1\,
dz_1^{\,\prime}\, dz_2\, dz_2^{\,\prime} \times \nonumber
\\
&&\qquad\qquad \phi_0^{(c.m.)}(\hbox{$\frac{z_1 +
z_2}{2}$})\,
{\phi_0^{(c.m.)}}^*\!(\hbox{$\frac{z_1^{\,\prime} +
z_2}{2}$}) \times \nonumber
\\
&&\qquad\qquad {\phi_0^{(c.m.)}}^*\!(\hbox{$\frac{z_1 +
z_2^{\,\prime}}{2}$})\,
\phi_0^{(c.m.)}(\hbox{$\frac{z_1^{\,\prime} +
z_2^{\,\prime}}{2}$}) \times \nonumber
\\
&&\qquad\qquad \phi^\mathrm{(rel)}_{n_{z_1}}(z_1 \!-\! z_2)\,
{\phi^\mathrm{(rel)}_{n_{z_2}}}^*(z_1^{\,\prime} \!-\! z_2) \times
\nonumber
\\
&&\qquad\qquad {\phi^\mathrm{(rel)}_{n_{z_3}}}^*(z_1 \!-\!
z_2^{\,\prime})\, \phi^\mathrm{(rel)}_{n_{z_4}}(z_1^{\,\prime}
\!-\! z_2^{\,\prime}). \label{int_J_def}
\end{eqnarray}

Using the expressions for the functions
$\Phi_{n,m}^{(c.m.)}$ and $\phi^\mathrm{(rel)}_{n_z}$ [Eqs.
(\ref{fd-states}) and (\ref{z_states}), where the parameters $\bar
\Omega$ and $\bar \omega_z$ are replaced, respectively, by $2\bar
\Omega$ and $2\bar \omega_z$ for the CM states and by $\bar
\Omega/2$ and $\bar \omega_z/2$ for the relative motion states],
one obtains  the following explicit forms for these integrals
\begin{eqnarray}
&&I(n_1,n_2,n_3,n_4;m) = \frac{1}{\pi^4 4^{|m|}}\nonumber
\\
&&\qquad \sqrt{\frac{n_1!\,n_2!\,n_3!\,n_4!}
{(n_1\!+\!|m|)!\,(n_2\!+\!|m|)!\,(n_3\!+\!|m|)!\,(n_4\!+\!|m|)!}}
\nonumber
\\[.5ex]
&&\qquad \int \cdots \int dx_1\, dy_1\, dx_1^{\,\prime}\,
dy_1^{\,\prime}\, dx_2\, dy_2\, dx_2^{\,\prime}\, dy_2^{\,\prime}
\\
&&\qquad\qquad e^{-(x_1^{\,2} + y_1^{\,2} + x_1^{\,\prime 2} +
y_1^{\,\prime 2} + x_2^{\,2} + y_2^{\,2} + x_2^{\,\prime 2} +
y_2^{\,\prime 2})} \nonumber
\\[-.5ex]
&& \qquad [(x_1 \!-\! x_2) \!+\! i(y_1 \!-\! y_2)]^{|m|}\,
[(x_1^{\,\prime} \!-\! x_2) \!-\! i(y_1^{\,\prime} \!-\!
y_2)]^{|m|} \nonumber
\\
&& \qquad [(x_1 \!-\! x_2^{\,\prime}) \!-\! i(y_1 \!-\!
y_2^{\,\prime})]^{|m|}\, [(x_1^{\,\prime} \!-\! x_2^{\,\prime})
\!+\! i(y_1^{\,\prime} \!-\! y_2^{\,\prime})]^{|m|} \nonumber
\\
&& \qquad \mathrm{L}_{n_1}^{|m|}(\hbox{$\frac{(x_1 - x_2)^2 + (y_1
- y_2)^2}{2}$})\,
\mathrm{L}_{n_2}^{|m|}(\hbox{$\frac{(x_1^{\,\prime} - x_2)^2 +
(y_1^{\,\prime} - y_2)^2}{2}$}) \nonumber
\\
&& \qquad \mathrm{L}_{n_3}^{|m|}(\hbox{$\frac{(x_1 -
x_2^{\,\prime})^2 + (y_1 - y_2^{\,\prime})^2}{2}$})\,
\mathrm{L}_{n_4}^{|m|}(\hbox{$\frac{(x_1^{\,\prime} -
x_2^{\,\prime})^2 + (y_1^{\,\prime} - y_2^{\,\prime})^2}{2}$}),
\nonumber \label{int_I}
\end{eqnarray}
\begin{eqnarray}
&&J(n_{z_1},n_{z_2},n_{z_3},n_{z_4}) = \nonumber
\\
&&\qquad \frac{1}{\pi^2
\sqrt{2^{n_{z_1}+n_{z_2}+n_{z_3}+n_{z_4}}\, n_{z_1}!\, n_{z_2}!\,
n_{z_3}!\, n_{z_4}}} \nonumber
\\[.5ex]
&&\qquad \int \cdots \int dz_1\, dz_1^{\,\prime}\, dz_2\,
dz_2^{\,\prime}\, e^{-(z_1^{\,2} + z_1^{\,\prime 2} + z_2^{\,2} +
z_2^{\,\prime 2})}
\\[-.5ex]
&& \mathrm{H}_{n_{z_1}}\!(\hbox{$\frac{z_1 - z_2}{\sqrt{2}}$})\,
\mathrm{H}_{n_{z_2}}\!(\hbox{$\frac{z_1^{\,\prime} -
z_2}{\sqrt{2}}$})\, \mathrm{H}_{n_{z_3}}\!(\hbox{$\frac{z_1 -
z_2^{\,\prime}}{\sqrt{2}}$})\,
\mathrm{H}_{n_{z_4}}\!(\hbox{$\frac{z_1^{\,\prime} -
z_2^{\,\prime}}{\sqrt{2}}$}). \nonumber \label{int_J}
\end{eqnarray}
 Evidently, the $I$ and $J$ integrals do not depend on the
parameters $\bar\Omega$ and $\bar{\omega}_z$.

Some properties of the $I$ and $J$ integrals are
\renewcommand{\theenumi}{(\roman{enumi})}
\begin{enumerate}
\item $I(n_1,n_2,n_3,n_4;m) = 0$ when $n_1 + n_4 \neq n_2 + n_3$;
\item if $(n_1^\prime,n_2^\prime,n_3^\prime,n_4^\prime)$ is a
permutation of indices $(n_1,n_2,n_3,n_4)$ such that $n_1 + n_4 =
n_2 + n_3$ and  $n_1^\prime + n_4^\prime = n_2^\prime +
n_3^\prime$, then
$I(n_1^\prime,n_2^\prime,n_3^\prime,n_4^\prime;m) =
I(n_1,n_2,n_3,n_4;m)$;
\item $I(n_1,n_2,n_3,n_4;0) = J^2(n_1,n_2,n_3,n_4)$.
\end{enumerate}

\section{Relations between the IP and CM basis elements}
\label{sec:basis-trans}

At $n_{c.m.} = n = 0$ and $m_{c.m.}, m \ge 0$,
the product of the Fock-Darwin states for the c.m. and relative
motions  (given by Eqs.~(\ref{cmfd-states}) and
(\ref{relfd-states})) is
\begin{eqnarray}
&&\Phi^{(c.m.)}_{0,m_{c.m.}}(\mathbf{R})
\Phi^\mathrm{(rel)}_{0,m}(\mathbf{r}_{12}) = 
\nonumber\\
&&\qquad 
{\sqrt\frac{(2\bar{\Omega})^{m_{c.m.}+1}}
{\pi\,m_{c.m.}!}}\, R^{m_{c.m.}} e^{-\bar{\Omega}R^2}
e^{i m_{c.m.}\varphi_{c.m.}} \times 
\nonumber\\
&&\qquad \sqrt{\frac{(\bar{\Omega}/2)^{m+1}}{\pi\,m!}}\,
r_{12}^m\, e^{-\frac{1}{4}\bar{\Omega}r_{12}^2} e^{im\varphi_{12}}
= \nonumber
\\
&&\qquad \frac{\bar{\Omega}}{\pi}
\sqrt{\frac{2^{m_{c.m.}-m}\, \bar{\Omega}^M}{m!\,
m_{{c.m.}}!}}\, e^{-\bar{\Omega}(R^2+r_{12}^2/4)} \times
\nonumber
\\
&&\qquad (R\,e^{i\varphi_{c.m.}})^{m_{c.m.}} (r_{12}\,
e^{i\varphi_{12}})^m, \label{CM-product}
\end{eqnarray}
where $M = m_1+m_2 = m_{c.m.}+m$. After the transition to the
individual particle coordinates, applying the relation
$2R^2+r_{12}^2/2 = r_1^2 + r_2^2$ and the binomial expansions
\begin{eqnarray}
&&(R\,e^{i\varphi_{c.m.}})^{m_{c.m.}} =
\bigg(\frac{r_1\,e^{i\varphi_1} +
r_2\,e^{i\varphi_2}}{2}\bigg)^{m_{c.m.}} = \nonumber
\\
&&\frac{1}{2^{m_{c.m.}}} \sum_{j=0}^{m_{c.m.}} \bigg(
\begin{matrix}
m_{c.m.} \\ j
\end{matrix}
\bigg) r_1^{m_{c.m.}-j}\times\nonumber\\
&&\times r_2^j\,
e^{i(m_{c.m.}-j)\varphi_1}\, e^{ij\varphi_2}\,,
\end{eqnarray}
taking into account
\begin{eqnarray}
&&(r_{12}\,e^{i\varphi_{12}})^m = (r_1\,e^{i\varphi_1} -
r_2\,e^{i\varphi_2})^m = \nonumber
\\
&& \sum_{k=0}^m \bigg(
\begin{matrix}
m \\ k
\end{matrix}
\bigg) r_1^{m-k}\, r_2^k\, e^{i(m-k)\varphi_2}\, e^{ik\varphi_2},
\end{eqnarray}
the product (\ref{CM-product}) transforms to
\begin{eqnarray}
&&\Phi^{(c.m.)}_{0,m_{c.m.}}(\mathbf{R})
\Phi^\mathrm{(rel)}_{0,m}(\mathbf{r}_{12}) = \nonumber
\\
&&\qquad \frac{\bar\Omega}{\pi}
\frac{{\bar\Omega}^{M/2}}{\sqrt{2^M\, m_{c.m.}!\, m!}}\,
e^{-\frac{1}{2}\bar\Omega(r_1^2+r_2^2)} \times \nonumber
\\
&&\qquad \sum_{j=0}^{m_{c.m.}} \sum_{k=0}^m (-1)^k \bigg(
\begin{matrix}
m_{c.m.} \\ j
\end{matrix}
\bigg) \bigg(
\begin{matrix}
m \\ k
\end{matrix}
\bigg) r_1^{M-j-k}\, r_2^{j+k} \times \nonumber
\\
&&\qquad\qquad e^{i(M-j-k)\varphi_1}\, e^{i(j+k)\varphi_2}.
\end{eqnarray}
Applying, finally, the expression (\ref{fd-states}) for the
Fock-Darwin states, now for the individual particles, the product
(\ref{CM-product}) takes the form
\begin{eqnarray}
&&\Phi^{(c.m.)}_{0,m_{c.m.}}(\mathbf{R})
\Phi^\mathrm{(rel)}_{0,m}(\mathbf{r}_{12}) = \nonumber
\\
&& \sum_{j=0}^{m_{c.m.}} \sum_{k=0}^m (-1)^k \bigg(
\begin{matrix}
m_{c.m.} \\ j
\end{matrix}\bigg)
\bigg(
\begin{matrix}
m \\ k
\end{matrix}
\bigg) \sqrt{\frac{(M\!-\!j-\!k)!\,(j+k)!}{2^M\, m_{c.m.}!\,
m!}} \times \nonumber
\\
&&\qquad\qquad \Phi_{0,M-j-k}(\mathbf{r}_1)
\Phi_{0,j+k}(\mathbf{r}_2).
\end{eqnarray}

Analogously we derive the inverse transformation
\begin{eqnarray}
&&\Phi_{0,m_1}(\mathbf{r}_1) \Phi_{0,m_2}(\mathbf{r}_2) =
\nonumber
\\
&&\sum_{j=0}^{m_1} \sum_{k=0}^{m_2} (-1)^k \bigg(
\begin{matrix}
m_1 \\ j
\end{matrix}
\bigg) \bigg(
\begin{matrix}
m_2 \\ k
\end{matrix}
\bigg) \sqrt{\frac{(M\!-\!j-\!k)!\,(j+k)!}{2^M\, m_1!\, m_2!}}
\times \nonumber
\\
&&\qquad\qquad\Phi^{(c.m.)}_{0,M-j-k}(\mathbf{R})
\Phi^\mathrm{(rel)}_{0,j+k}(\mathbf{r}_{12}).
\end{eqnarray}

\section{An approximate expression for $\langle\rho_{12}^2\rangle$ in the ground state}
\label{sec:meanv-rho2}

The mean value of $\rho_{12}^2$ in an arbitrary state
$\psi_\mathrm{rel}$, given in the form of expansion (17), is
\begin{eqnarray}
\langle\rho_{12}^2\rangle = \sum_{n,n^\prime} \sum_{n_z}
{b_{n,n_z}^{(m)}}^* b_{n^\prime,n_z}^{(m)} \langle
n,m|\rho_{12}^2|n^\prime,m \rangle\,.
\end{eqnarray}
Here, the matrix elements
\begin{eqnarray}
\langle n,m|\rho_{12}^2|n^\prime,m \rangle &=&
\frac{\hbar}{\mu\Omega} [ (2n + m + 1)\, \delta_{n,n^\prime} -
\nonumber
\\
&&\sqrt{n(n+m)}\, \delta_{n-1,n^\prime} -
\\[.5ex]
&&\sqrt{(n+1)(n+m+1)}\, \delta_{n+1,n^\prime}] \nonumber
\end{eqnarray}
are  calculated between the Fock-Darwin states
$\Phi_{n,m}^\mathrm{(rel)}$. The Kronecker symbols cancel the sum
via $n^\prime$, and, as result, one has
\begin{eqnarray}
\label{gro}
\langle\rho_{12}^2\rangle &=& \frac{\hbar}{\mu\Omega} \sum_{n,n_z}
\Big[\,|b_{n,n_z}^{(m)}|^2 (2n + m + 1) - \nonumber
\\
&& ({b_{n+1,n_z}^{(m)}}^{\!\!\!\!*} b_{n,n_z}^{(m)} +
{b_{n,n_z}^{(m)}}^{\!\!*} b_{n+1,n_z}^{(m)}) \times
\\[.5ex]
&& \qquad \sqrt{(n+1)(n+m+1)}\,\Big]. \nonumber
\end{eqnarray}

For a typical two-electron QD in the ground state,
the relation  $|b_{0,0}^{(m)}|
\gg |b_{1,0}^{(m)}| \gg |b_{0,1}^{(m)}|, |b_{2,0}^{(m)}|,\ldots$
holds.
Keeping only the terms with $b_{0,0}^{(m)}$ and $b_{1,0}^{(m)}$
and assuming that $|b_{0,0}^{(m)}|^2 + |b_{1,0}^{(m)}|^2 \approx
1$, the general expression (\ref{gro}) is reduced to the form
\begin{eqnarray}
\langle\rho_{12}^2\rangle &=& \frac{\hbar}{\mu\Omega} \Big[\,m + 1
+ 2|b_{1,0}^{(m)}|^2 - \nonumber
\\
&& ({b_{1,0}^{(m)}}^* b_{0,0}^{(m)} + {b_{0,0}^{(m)}}^*
b_{1,0}^{(m)}) \sqrt{m+1}\,\Big].
\end{eqnarray}
Finally, since the $b$-coefficients in the expansion of
$\psi_\mathrm{rel}$ with a fixed value of $m$ can  always be chosen
to be real, one obtains
\begin{equation}
\langle\rho_{12}^2\rangle = \frac{\hbar}{\mu\Omega} \Big[\,m + 1 -
2 {b_{0,0}^{(m)}} b_{1,0}^{(m)} \sqrt{m+1} + 2{b_{1,0}^{(m)}}^2\,
\Big].
\end{equation}

\end{document}